\documentclass[nofootinbib,aps,preprint]{revtex4}
\usepackage[dvips]{graphicx}
\usepackage{graphicx}
\usepackage{amsmath}

\usepackage{axodraw}
\usepackage{epsfig}
\usepackage[dvips]{graphics,color}

\def\be{\begin{equation}}
\def\ee{\end{equation}}
\def\bea{\begin{eqnarray}}
\def\eea{\end{eqnarray}}
\def\bc{\begin{center}}
\def\ec{\end{center}}
\def\notder{{\not\! \partial}}
\def\notcder{{\not\!\! D}}
\def\notA{{\not\!\! A}}
\def\notW{{\not\!\! W}}
\def\notZ{{\not\! Z}}
\def\notB{{\not\!\! B}}
\def\noteq{{\not=}}

\begin{document}

\title{Weak Interactions in Atoms and Nuclei: The Standard Model and Beyond}
\author{M.J. Ramsey-Musolf} 
\affiliation{Kellogg Radiation Laboratory, Caltech, Pasadena, CA 91125}
\affiliation{Department of Physics, University of Connecticut, Storrs, CT 06269}
\author{J. Secrest}
\affiliation{Department of Physics, College of William and Mary, Williambsburg, VA 23187}
\begin{abstract}
Studies in nuclear and atomic physics have played an important role in
developing our understanding of the Standard Model of electroweak interactions.
We review the basic ingredients of the Standard Model, and discuss some key
nuclear and atomic physics experiments used in testing these ideas. We also
summarize the conceptual issues of the Standard Model that motivate the
search for new physics.
\end{abstract}
\maketitle
\pagenumbering{arabic}
\section{Introduction}
The quest for a unified description of all known forces of nature is
something of a \lq\lq holy grail" for
physicists. At present, we possess a partial description, known as the
Standard Model\cite{smrefs}. In a nutshell, the
Standard Model (SM) is a unified gauge theory of the strong, weak, and
electromagnetic interactions, the content
of which is summarized by the group structure
\begin{equation}
\label{eq:smgroup}
{\mbox{SU(3)}}_C\times{\mbox{SU(2)}}_L\times{\mbox{U(1)}}_Y \ \ \ ,
\end{equation}
where the first factor refers to the theory of strong interactions, or
Quantum Chromodynamics (QCD), and the
latter two factors describe the theory of electroweak interactions.
Although the theory remains incomplete, its
development represents a triumph for modern physics. Historically, nuclear
and atomic physics have played an
important role in uncovering the structure of the strong and electroweak
interactions. In this series of
lectures, I will attempt to give you some sense of that history, as well as
some feel for the parts being played
by nuclear and atomic physics in looking for physics beyond the SM. I will
focus on the electroweak sector of the
theory, though you should keep in mind that studies of QCD constitute one
of the primary thrusts of nuclear
physics today.

Before delving into the details of the SM, it is important to appreciate
just what an achievement Eq.
(\ref{eq:smgroup}) represents.  One way to do this is to consider some of
the basic properties off the four
forces of nature.
\begin{center}
\begin{tabular}{|c|c|c|c|c|} \hline
Interaction	& Range(fm)	&  Strength	& TimeScale &$\sigma$($\mu$b)  \\ \hline
		                  \hline
Gravity		& $\infty$			& $G_{N}M^{2}_{p}$ &$\times$	& $\times$ \\ \hline

Weak 		& $10^{-3}$	 & $G_{F}M^{2}_{p}$ & $\geq 10^{-8}$ & $10^{-8}$ \\ \hline

Strong		& $\leq 1$	 & $\alpha_{S}(r)$ &$10^{-23}$  &$10^{4}$ \\ \hline

EM         	& $\infty$	 & $\alpha$      & $10^{-20}$ &10	 \ \\ \hline
\end{tabular}
\end{center}

In addition to this seemingly disparate set of characteristics, each of the
forces has some unique features of
its own.  For example, the gravitational and EM interactions get weaker the
farther apart the given ``charges''
are, whereas the strong interaction behaves just the opposite, as the
distance $r$ increases, $\alpha_{S}$ grows.
This feature is related to the fact that one never sees individual unbound
quarks and that strong interaction effects at
distance scales at about 1 fm are hard to compute.\\
Another set of unique features has to do with how each interaction obeys
different ``discrete'' symmetries:
\begin{center}
\begin{tabular}{|c|c|c|c|c|}\hline
Parity	& P	& $\overrightarrow{x} \rightarrow \overrightarrow{-x}$  &
$t \rightarrow t$ &$\overrightarrow{s}
\rightarrow \overrightarrow{s}$ \\  \hline
Time Reversal & T & $\overrightarrow{x} \rightarrow \overrightarrow{x}$ &
$t \rightarrow -t$ &
$\overrightarrow{s} \rightarrow \overrightarrow{s}$ \\  \hline
Charge Conjugation & C & $Q \rightarrow -Q$ & $e^{+} \rightarrow e^{-}$ &{
\em etc.} \\ \hline
\end{tabular}
\end{center}
A simple way to visualize these symmetries is as follows. A parity
transformation (P) involves an inversion of
a physical system through the origin of co-ordinates. This inversion can be
completed in two steps. First,
reflect the system in a mirror. Second, rotate the reflected system by
$180^\circ$ about the normal to the mirror.
However, since fundamental interactions are rotationally invariant (angular
momentum is conserved), we may omit
the second step. Thus, a parity transformation is essentially a mirror
reflection. Note that the \lq\lq
handedness" of a particle -- the relative orientation of its momentum and
spin -- reverses under mirror
reflection. A time reversal transformation (T) amounts first to thinking of
a physical process as being like
a movie. Under T, the movie is run backwards through the projector.
Finally, a charge conjugation
transformation (C) turns a particle into its antiparticle.

Here's how the different interactions rate with respect to these symmetries:
\begin{center}Symmetry ``Score Card''\\
\begin{tabular}{|c|c|c|c|}\hline
Symmetry	& Strong	& EM		& Weak \\ \hline
P		& yes	& yes	& no \\ \hline
C		& yes	& yes      & no \\ \hline
T		& yes	& yes	& no  \\ \hline
PCT  		& yes	& yes	& yes \\ \hline
\end{tabular}
\end{center}

Clearly, the weak interaction is a flagrant symmetry violator, whereas the
strong interaction and EM interactions
respect P,C, and T individually.  So how can it be that all three forces
fall into a unified model?\\
Given these differences, it is truly remarkable that physicists have
figured out how to describe three out of
four forces in a unified theory.  It's a highly non-trivial accomplishment.\\
Now lets look in detail at the essential elements of the electroweak part
of the Standard Model.

\section{Basic Ingredients of the Standard Model}
The essential building blocks of the electroweak sector of the Standard
Model are the following:\\
\medskip
\noindent
1. Gauge Symmetry $\Longleftrightarrow$ gauge bosons, parity violation\\
2. Representations $\Longleftrightarrow$ bosons and quarks\\
3. Family Replication, mixing and universality $\Longleftrightarrow$ CP
Violation\\
4. Spontaneous Symmetry Breaking $\Longleftrightarrow$ Higgs Boson,
$M_{Z}$, $M_{W} \not{=} 0 $\\

\subsection{Gauge Symmetry}
Lets start with the more familiar case of the electromagnetism and see how
the properties can all be derived from the principle of gauge invariance.\\
\noindent
Consider the Dirac equation for a free electron:
\begin{equation}
\label{eq:dirac1}
(i \notder - m)\psi = 0\ \ \ .
\end{equation}
Suppose we now make the local transformation:
\begin{equation}
\psi(x) \rightarrow \psi(x)e^{i\alpha(x)}=\psi^{\prime}(x)\ \ \ .
\label{eq:gauge1}
\end{equation}
This is called a U(1) transformation.\\
The Dirac equation is not invariant under this transformation. If
$\psi^{\prime}$ satisfies (\ref{eq:dirac1}) then one
has:
\begin{equation}
(i \notder - m)e^{i\alpha(x)}\psi =(i \notder - m)\psi - \psi \notder \alpha= 0\ \ \ .
\label{eq:dirac2}
\end{equation}
To make (\ref{eq:dirac1}) invariant under (\ref{eq:gauge1}), one can
replace $\partial_{\mu}$ by the covariant
derivative $D_{\mu}$:
\begin{equation}
\label{eq:covder}
D_{\mu} = \partial_{\mu} -ieA_{\mu}\ \ \  ,
\end{equation}
where $A_{\mu}$ is a gauge field identified with the photon.  We require
$A_{\mu}$ to transform in
such away as to remove the unwanted term in (\ref{eq:dirac2}).
\begin{equation}
\label{eq:gauge2}
A_{\mu} \rightarrow A_{\mu}^{\prime} = A_{\mu} + {1 \over e } \partial_{\mu}\alpha
\end{equation}
when $\psi \rightarrow \psi^{\prime}=e^{i\alpha}\psi$\\
Thus, we obtain a new Dirac equation:
\begin{equation}
(i\notcder - m)\psi =0 \ \ \  .
\end{equation}
Under a gauge transformation, one has
\begin{eqnarray}
 (i\notcder^{\prime} - m)\psi^{\prime} &=&0\ \ \longrightarrow\nonumber \\
=[i(\notder-ie\notA^{\prime}_{\mu})-m]\psi^{\prime}
&=&[i(\notder-ie\notA_{\mu}-i\notder\alpha)-m]\psi\nonumber \\
=e^{i\alpha}[i(\notder-ie\notA)-m]\psi &+&
e^{i\alpha}(\notder\alpha)\psi-e^{i\alpha}(\notder\alpha)\psi\nonumber \\
&=&e^{i\alpha}(i\notcder - m)\psi=0\ \ \ .
\end{eqnarray}
Cancelling through the $e^{i\alpha}$ yields
\begin{equation}
(i\notcder - m)\psi =0\ \ \  {\hbox{\rm if}} \ \ \ (i\notcder^{\prime} -m)\psi^{\prime} =0\ \ \ .
\end{equation}
One recognizes the replacement of $\partial_{\mu} \rightarrow
D_{\mu}=\partial_\mu-ieA_{\mu}$ as the usual minimal
substitution that gives us the interaction of the electrons with the vector
potential in quantum mechanics.  Apparently,
requiring the Dirac equation to be invariant under U(1) gauge
transformation (\ref{eq:gauge1}) and
(\ref{eq:gauge2}) leads to  a familiar result from quantum mechanics.

This idea can be generalized to other symmetry transformations.  A simple
generalization is a group of
transformations called SU(2).  Let's  define the following matrices:
\begin{eqnarray}
\tau_{1} = \begin{pmatrix} 0 & 1 \\ 1 &  0 \end{pmatrix}\ \ \
 \tau_{2}=\begin{pmatrix} 0 & -i \\ i & 0 \end{pmatrix} \ \ \
\tau_{3}=\begin{pmatrix}1 & 0 \\ 0 & -1 \end{pmatrix}\ \ \ .  \nonumber \\
\end{eqnarray}
They satisfy:
\begin{eqnarray}
\left[{\tau_i \over 2},{\tau_j \over 2}\right] &=& i\epsilon_{ijk}{\tau_k \over 2}  \nonumber\\
\left[{\tau_i \over 2},{\tau_j \over 2}\right]_{+} &=& {1 \over 2}\delta_{ij}  \nonumber \\
{\rm Tr}\left({\tau_i \over 2} {\tau_j \over 2}\right) &=& {1 \over 2}\delta_{ij}\ \ \ . \nonumber
\end{eqnarray}
One defines a group of transformations -- called SU(2) --  whose elements
are:
\begin{equation}
\label{eq:gauge3}
U({\vec\alpha})=e^{i{\vec\alpha(x)}\cdot {{\vec\tau} \over 2}}
\end{equation}
where
\begin{eqnarray}
 \nonumber
{\vec\alpha(x)}&=&(\alpha_{1}(x),\alpha_{2}(x),\alpha_{3}(x))\\
 \nonumber
{\vec\tau}&=&(\tau_{1},\tau_{2},\tau_{3})\ \ \ ,
\end{eqnarray}
with the $\alpha_{i}$'s being a continuously varying function of $x^{\mu} =
(t,{\vec x}$).  The transformations
$U(\overrightarrow{\alpha})$ act on a two component vector.  For example, let
\begin{equation}
\Psi_{l} =\begin{pmatrix}\psi_{\nu}(x) \\ \psi_{e}(x) \end{pmatrix}
\end{equation}
denote a lepton wavefunction\footnote{Here, $\psi_{\nu, e}$ are
four-component Dirac spinors.}. Then
under the transformation (\ref{eq:gauge3})
\begin{equation}
\label{eq:su2a}
\Psi_{l} \rightarrow \Psi_{l}^{\prime}=U({\vec\alpha})\Psi_{l}
\end{equation}
In order for the Dirac equation to be invariant under (\ref{eq:su2a}), one
must define a new gauge covariant derivative for SU(2):
\begin{equation}
D_{\mu}=\partial_{\mu}-ig{\tau \over 2} \cdot {\vec W_{\mu}}\ \ \ ,
\end{equation}
where $g$ is an SU(2) analog of electric charge and
\begin{equation}
{\vec W_{\mu}} =(W_{\mu}^{1},W_{\mu}^{2},W_{\mu}^{3})
\end{equation}
is a set of 3 fields analogous to $A_{\mu}$.  It
turns out that if ${\vec W}_\mu$ transforms as
\begin{equation}
\label{eq:su2b}
W_{\mu}^{i} \rightarrow  W_{\mu}^{i\prime}=W_{\mu}^{i}-\epsilon^{ijk}\alpha_{j}W^{k}_{\mu}-{1 \over g}\partial_{\mu}\alpha_{i}
\end{equation}
when
\begin{equation}
\Psi_{l} \rightarrow U({\vec\alpha})\Psi_{l} = \Psi_{l}^{\prime}
\end{equation}
then the Dirac equation for $\psi_{\nu}$ and $\psi_{e}$  will be invariant
under the transformations
(\ref{eq:su2a}) and (\ref{eq:su2b}):
\begin{equation}
\label{eq:su2dirac}
(i\notder+g{\overrightarrow{\tau} \over 2} \cdot
{\overrightarrow{\notW_{\mu}}} - \hat{M})\Psi_{l}=0\ \ \ .
\end{equation}
Note that the extra term in the transformation law
for $W^{i}_{\nu}$ -- the $\epsilon^{ijk}\alpha_{j}W^{k}_{\nu}$ term -- is a
consequence of the fact that this
group of transformations is non-Abelian, that is $\tau_{i}$ and $\tau_{j}$
do not commute.\\
Note also that in Eq. (\ref{eq:su2dirac}) we have introduced a mass term
($\hat{M}$), whose origins I
will discuss later.

\begin{figure}[h]
\psfig{figure=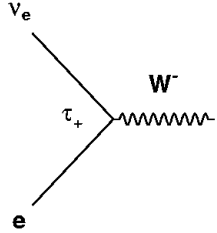,height=1.4in}
\caption{Charge raising weak interaction current.}
\label{feyn:diag1}
\end{figure}

As in the case of the U(1)$_{EM}$ gauge field $A_\mu$, which we corresponds
to the photon, the SU(2) gauge fields
$W^{i}_{\nu}$ should also be associated with spin-one particles. To
identify the character of these particles,
first define:
\bc
$$\tau_{+}={1 \over 2}(\tau_{1} + i\tau_{2}) = \begin{pmatrix} 0 & 1 \\ 0 &
0\end{pmatrix}$$\\
$$\tau_{-}={1 \over 2}(\tau_{1} - i\tau_{2}) = \begin{pmatrix} 0 & 0 \\ 1 &
0 \end{pmatrix}$$\\
\ec
and $W^{\pm}_{\mu}={1 \over \sqrt{2}}(W^{1}_{\mu} \pm i W^{2}_{\mu})$.
Then
\be
{\vec\tau} \cdot {\vec W_{\mu}} = \sqrt{2}(\tau_{+}W^{-}_{\mu} +
\tau_{-}W^{+}_{\mu})+\tau_{3}W^{3}_{\mu}\ \ \ .
\end{equation}
Now
\be
\tau_{+}\Psi_{l} = \begin{pmatrix} 0 & 1 \\ 0 & 0 \end{pmatrix}\begin{pmatrix}
\psi_{\nu} \\ \psi_{e} \end{pmatrix} = \begin{pmatrix} \psi_{e} \\ 0
\end{pmatrix}\ \ \ .
\end{equation}
In short, acting with $\tau_+$ on the lepton wavefunction transforms an
electron wavefunction
into one for a neutrino:
$\psi_{e} \rightarrow \psi_{\nu}$. Similarly,
\be
\tau_{-}\Psi_{l} = \begin{pmatrix} 0 & 0 \\ 1 & 0
\end{pmatrix}\begin{pmatrix} \psi_{\nu} \\  \psi_{e}
 \end{pmatrix}
=\begin{pmatrix} 0 \\ \psi_{\nu} \end{pmatrix}\ \ \ .
\end{equation}
This turns a neutrino into an electron: $\psi_{\nu} \rightarrow \psi_{e}$.
One can represent the action of ${\vec\tau} \cdot {\vec W_{\mu}}$ on
$\Psi_{l}$
in Fig.  \ref{feyn:diag1}:
\bc
\ec
From the fact that weak interactions turn  $e \leftrightarrow \nu$,
one is lead to identify the $W^{\pm}_{\mu}$ fields with the $W^{\pm}$
particles.\\
It would be tempting to identify the $W^{3}_{\mu}$ with the $Z^0$ boson.
However, it turns out to be impossible to do that and end up with the right
masses for the $W^{\pm}$and $Z^0$
bosons.  Moreover, since there exists another neutral boson -- the massless
photon, $\gamma$ --  one needs two
neutral fields to make the
$Z^0$  and $\gamma$.  A nice way to produce both bosons is to {\em mix} the
$W^{3}_{\mu}$ with another
gauge boson -- called $B_{\mu}$ -- that transforms as a singlet under SU(2)
transformations. In short,
$(W_\mu^3, B_\mu)\to (A_\mu, Z_\mu)$. \\
Before fleshing this idea out, however, we need to revisit the parity
transformation discussed earlier. To that end,
let's define:
\begin{eqnarray}
\psi_{L} &=& {1 \over 2}(1-\gamma_{5})\psi\\
\psi_{R} &=& {1 \over 2}(1+\gamma_{5})\psi\ \ \ .
\end{eqnarray}
One can show that if $\psi$ is massless, then  $\psi_{L}$ always has
negative helicity and $\psi_{R}$ always has positive helicity.\\
In short:
\bc
$\psi_{L}:\ \ \ $	$h=\hat{s} \cdot \hat{p}= -1$\\
$\psi_{R}:\ \ \ $	$h=\hat{s} \cdot \hat{p}= +1$\\
\ec
\noindent Under a parity transformation, $\hat{s} \cdot \hat{p}$ changes
sign which is equivalent to saying that $\psi_{L}
\leftrightarrow\psi_{R}$.\\
One knows that processes like $\mu$-decay or $\beta$-decay,
the neutrinos always come with helicity $h$=-1, which implies parity
symmetry is broken.  Only left handed
particles participate in these types of weak interactions.\\
Since the interactions of $W^{\pm}_{\mu}$ with leptons arises in  Eq.
(\ref{eq:su2dirac}) because of SU(2)
invariance,  and one needs only left handed particles to have interactions
with the $W^{\pm}_{\mu}$,
one must
modify the transformations rule:
\begin{eqnarray}
\nonumber
{\mbox{SU(2)}} & \rightarrow & {\mbox{SU(2)}}_{L}\\
\nonumber
\Psi^{L}_{l} &\rightarrow& U({\vec\alpha})\Psi^{L}_{l}
\ \ \ {\mbox{(doublet)}}\\
\nonumber
\psi_{e_{R}} &\rightarrow&  \psi_{e_{R}}  \ \ \ {\mbox{(singlet)}}\ \ \ .
\end{eqnarray}
In short, only left-handed particles undergo SU(2) transformations, while
right-handed particles are unaffected.
Thus, the Dirac equation for $  \psi_{e_{R}} = {1 \over
2}(1+\gamma_{5})\psi_{e}$ is unchanged under SU(2)$_{L}$
so there is no need for the $\overrightarrow{\tau} \cdot
\overrightarrow{W_{\mu}}$ term to maintain invariance.
Hence, one has the SU(2)$_{L}$ in $SU(2)_{L} \times U(1)_{Y}$. Note that we
have not included a transformation $\psi_{\nu_R}$ in the list of
transformations. The reason is that
prior to the discovery of neutrino oscillations, right-handed neutrinos
were not observed to participate in
low-energy weak interactions. \\
Now back to the $Z^0$ and $\gamma$.  Nature gives us four bosons in the
electroweak interaction:
two charged particles ($W^{\pm}$) and two neutral particles ($Z^0$,
$\gamma$).  So far, we have
identified two charged and one neutral gauge boson. We need to include one
more neutral gauge boson
having no accompanying particles. The way to accomplish this is to
introduce one more set of U(1)
transformations:
\bea
\psi \rightarrow e^{i\alpha(x)} \psi &=& \psi^{\prime}\nonumber \\
\partial_{\mu} &\rightarrow &\partial_{\mu} - ig^{\prime}{Y \over 2}
B_{\mu}\nonumber \\
B_{\mu} \rightarrow B_{\mu}^{\prime} &= & B_{\mu} + {1 \over g^{\prime}}
\partial_{\mu}\alpha\ \ \ .\nonumber
\eea
Here Y is called {\em hypercharge} (to distinguish it from the general EM
charge).
Now the Dirac equation for electrons and neutrinos is the following:
\bea
\label{eq:diracleft}
(i\notder + {g \over 2} \overrightarrow{\tau} \cdot \overrightarrow{\notW} + {g^{\prime} \over 2}Y\notB -\hat{M})\Psi^{L}_{l}&=&0\\
\label{eq:diracright}
(i\not{\partial} + {g^{\prime} \over 2}Y {\notB}-m_e)\psi^{R}_{e}&=&0\ \  \ .
\eea
Rewriting (\ref{eq:diracleft}) slightly:
\begin{eqnarray}
[i\not{\partial} &+& {g \over \sqrt{2}}(\tau_{+}\notW_{-} +
\tau_{-}\notW_{+})\nonumber \\
\label{eq:diracleft2}
&+& {g \over 2}\tau_{3}\notW_{3} + {g^{\prime} \over 2}Y\notB -\hat{M}]\Psi^{L}_{l}=0\ \ \ .
\end{eqnarray}
Now to get EM interactions for both the $e_{R}$ and $e_{L}$
and to make sure the $W^{\pm}$ and $Z^0$ are different in mass, one needs
the $Z^0$ and
$\gamma$ to be linear combinations of the $W_{3}$ and $B$.  One can
accomplish this by utilizing a unitary
transformation:
\begin{equation}
\label{eq:mixing}
\begin{pmatrix} Z^0_{\mu} \\ A_{\mu} \end{pmatrix} =
\begin{pmatrix} \cos\theta_{W} & -\sin\theta_{W}\\  \sin\theta_{W} &
\cos\theta_{W} \\
\end{pmatrix} \begin{pmatrix}
W^{3}_{\mu} \\ B_{\mu} \\ \end{pmatrix}\ \ \ .
\end{equation}
The angle $\theta_{W}$ is a supremely important parameter in the
electroweak standard model.
It is called the \lq\lq Weinberg angle", or \lq\lq weak mixing angle".
Inverting (\ref{eq:mixing}), one can substitute into
(\ref{eq:diracleft}) and (\ref{eq:diracright}) for $W^{3}_{\mu}$ and
$B_{\mu}$ to obtain:\\
Left-handed leptons:
\begin{eqnarray}
\nonumber
[i\notder &+& {g \over \sqrt{2}}(\tau_{+}\notW_{-} + \tau_{-}\notW_{+})\\
\nonumber
&+& (g \sin\theta_{W} {\tau_{3} \over 2} + g^{\prime} \cos\theta_{W} {Y \over 2})\notA\\
&+& (g \cos\theta_{W} {\tau_{3} \over 2} - g^{\prime} \sin\theta_{W} {Y \over 2})\notZ^0-{\hat M}]\Psi^{L}_{l}=0\label{eq:diracleft2}
\end{eqnarray}
Right-handed leptons: 
\begin{eqnarray}
[i\notder +(g^{\prime}\cos\theta_{W} {Y \over 2})\notA -(g^{\prime}
\sin\theta_{W} {Y \over
2})\notZ^0 -m_e]\psi_{e_{R}} = 0\label{eq:diracright2}
\end{eqnarray}
In order to restore the original EM gauge transformation law, one needs to
identify:
\begin{eqnarray}
g^{\prime} \cos\theta_{W} {Y_{R} \over 2} &=& eQ\\
g \sin\theta_{W} {\tau_{3} \over 2} + g^{\prime} \cos\theta_{W} {Y_L \over 2}&=&eQ\ \ \ .
\end{eqnarray}
This works if one takes:
\bea
g^{\prime}\cos\theta_{W} &=& g \sin\theta_{W}=e\nonumber \\
\label{eq:emcharge}
Y_{R}&=&2Q\\
Y_{L}&=&2(Q - T^{L}_{3})\nonumber
\eea
where $T^{L}_{3}\Psi^{L}_{l}=(\tau_{3}/ 2)\Psi^{L}_{l}$.\\
From Eqs. (\ref{eq:emcharge}) one also has
\begin{equation}
\sin \theta_{W} = {e \over g}\ \ \ {\mbox{and}}\ \ \ \tan \theta_{W} =
{g^{\prime} \over g}\ \ \ .
\end{equation}
Thus, the SU(2)$_{L}$ and U(1)$_{EM}$ interactions depend on three
parameters:
\bc
$g$,$g^{\prime}$, and $\sin \theta_{W}$\\
or\\
$g$,$e$, and $\sin\theta_{W}$\\
{\em etc.}\\
\ec
Lastly, let's rewrite our $Z^0$ couplings by eliminating $g^{\prime}$ in
terms of $g$ and $\sin
\theta_{W}$:\\ For the right handed sector:
\be
-g^{\prime}{ Y_{R} \over 2 } = -g {\sin^{2} \theta_{W} \over \cos
\theta_{W} }Q \equiv {gQ_{R}^W \over \cos
\theta_{W}}
\end{equation}
For the left handed sector:
\bea
\nonumber
g \cos \theta_{W} {\tau_{3} \over 2} - g^{\prime} \sin \theta_{W} {Y_{L}
\over 2}&=& \\
\nonumber
g \cos \theta_{W} {\tau_{3} \over 2} - g {\sin^{2} \theta_{W} \over \cos
\theta_{W}}(Q -
{\tau_{3} \over 2}) &=& \\
\nonumber
{g \over \cos \theta_{W}}(T^{L}_{3} - \sin^{2} \theta_{W} Q)&\equiv& \\
{g \over \cos \theta_{W}} Q_{L}^W&&\ \ \ ,
\eea
where $Q_{L,R}^W$ denote the left- and right-handed \lq\lq weak charges" of
the leptons.
To summarize, then, we have:
\begin{equation}
[i \notder + e Q \notA + {g Q_{R}^W\over\cos\theta_{W}} \notZ-m_e]\psi_{e_{R}}=0
\end{equation}
\bea
[i \notder &+& e Q \notA + {g Q_{L}\over \cos \theta_{W}} \notZ \\
&+& {g \over \sqrt{2}} (
\tau_{+}\notW_{-} +\tau_{-}\notW_{+}) - {\hat M}]\Psi_{l}^{L}=0\ \ \ ,
\eea
where
\bea
\label{eq:sum1}
g&=&{e \over \sin \theta_{W}}\\
\label{eq:sum2}
Q_{R}^W&=& -\sin^{2} \theta_{W}Q\\
\label{eq:sum3}
Q_{L}^W&=&T^{L}_{3}-\sin^{2}\theta_{W}Q\ \ \ .
\eea
\subsection{Representations}
The assignment of different SU(2)$_{L}$ and U(1)$_{Y}$ transformation
properties
to left- and right-handed fermions is called a choice of representations.
Left-handed fermions are
assigned to a doublet representations of SU(2)$_{L}$; right-handed fermions
transform under the singlet
representation.
\bc
$$\begin{pmatrix} \nu_{L}\\ e_{L} \end{pmatrix}  \hspace{1cm}\nu_{R}
\hspace{1cm}e_{R}$$\\
doublet	\hspace{1cm}singlet \hspace{1cm}    singlet\\
\ec
Note that although the $\nu_R$ transforms as a singlet, it has no weak
interactions
according to Eq. (\ref{eq:sum2}) since its electromagnetic charge is zero.

One can make the same assignment for quarks:
\bc
$$\begin{pmatrix} u_{L}\\ d_{L} \end{pmatrix}  \hspace{1cm}u_{R}
\hspace{1cm}d_{R}$$\\
doublet	\hspace{1cm}singlet \hspace{1cm}    singlet\\
\ec
Note that unlike the the right-handed neutrinos, both right-handed quarks
have weak
interactions since $Q_{u} \noteq 0 $ and $Q_{d} \noteq 0 $\\

\subsection{Family Replication, Universality, and Mixing}
Of course electrons, neutrinos, up quarks, and down quarks are not
all the elementary leptons and quarks.  They constitute the first
generation.  The remaining are
assigned to the second and third generations:\\

Second:
\bc
$$\begin{pmatrix} \nu^{L}_{\mu} \\ \mu_{L} \end{pmatrix}
\hspace{1cm}\begin{pmatrix} c \\ s \end{pmatrix}
\hspace{1cm}\mu_{R},c_{R},s_{R}$$
\ec
Third:
\bc
$$\begin{pmatrix} \nu^{L}_{\tau} \\ \tau_{L} \end{pmatrix}
\hspace{1cm}\begin{pmatrix} t \\ b
\end{pmatrix}\hspace{1cm}\tau_{R},t_{R},b_{R}$$\\
\ec
This repeated pattern of assignment is called fermion family replication.

An important feature of the standard model is that for each family,
the structure of the interactions with gauge bosons is the same. In other
words, there
is a {\em family universality}.

Moreover, the overall strength of the charged current
interactions and neutral current interactions is set by the same parameter,
$g$.  This feature is known
as {\em charged current/neutral current universality}.

Now this nice pattern of universality gets somewhat clouded for quarks
because the
quark eigenstates of the weak interaction can different from the quark
eigenstates of the
mass operator. The origin of this effect has to do with how the electroweak
symmetry is broken,
as discussed below.  The
relationship between the two sets of eigenstates can be expressed by
letting each negative charged
quark weak eigenstate be written as a linear combination of the three mass
eigenstates:
\begin{equation}
\begin{pmatrix} d \\ s \\ b\end{pmatrix}_{\rm WEAK} = {\hat V}
\begin{pmatrix} d \\ s \\ b \end{pmatrix}_{\rm MASS}\ \ \ ,
\end{equation}
where ${\hat V}$ is a 3 $\times$ 3 matrix known as the
Cabbibo-Kobayashi-Maskawa (CKM) matrix\cite{ckmrefs}.
In general, it is unitary and parameterized by 3 angles and a phase.

It turns out that one  need  not to write a similar relation for the
positive quarks
for algebraic reasons that will not be discussed here.
To illustrate, then, the weak quark doublet of the first generation is:\\
\bc
$$\begin{pmatrix} u \\ \widetilde{d} \end{pmatrix}_L, \hspace{1cm}
\widetilde{d}=V_{11}d + V_{12}s + V_{13}b$$\\
\ec
with
\begin{equation}
|V_{11}|^2+|V_{12}|^2+|V_{13}|^2 = 1
\end{equation}
following from the unitarity of ${\hat V}$.

\begin{figure}[h]
\hspace{.25cm}\psfig{figure=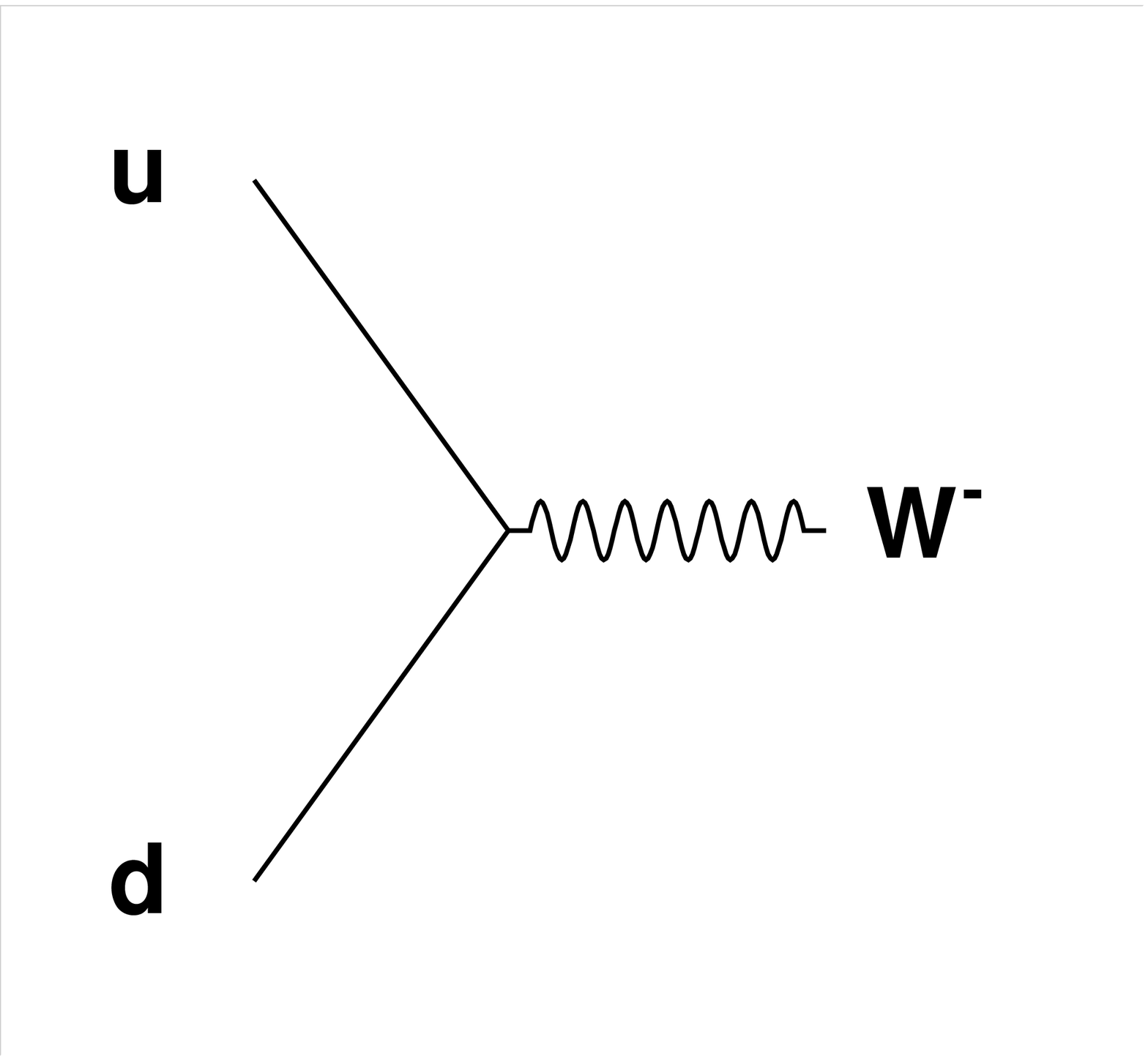,height=1.5in}
\psfig{figure=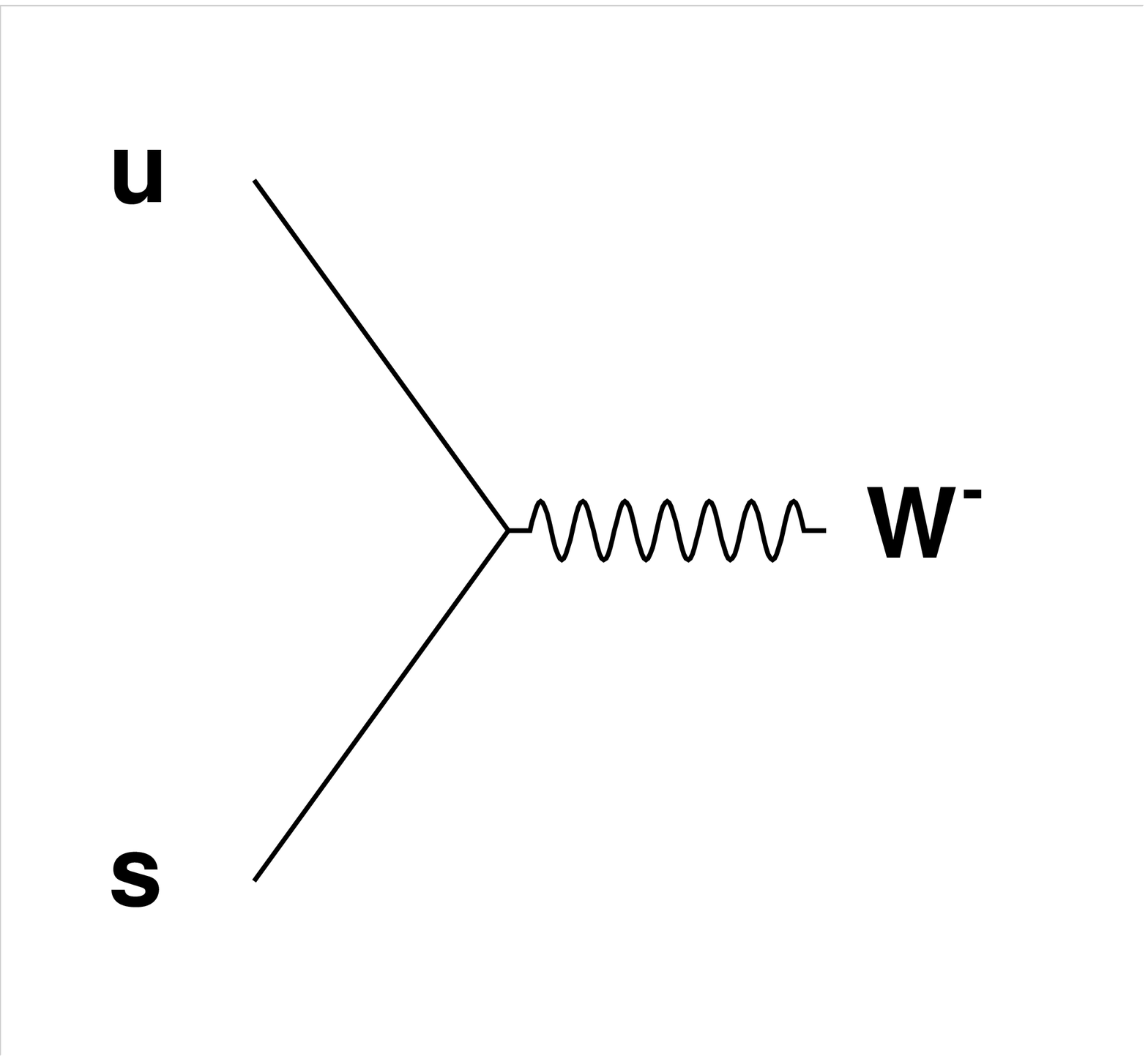,height=1.5in}
\psfig{figure=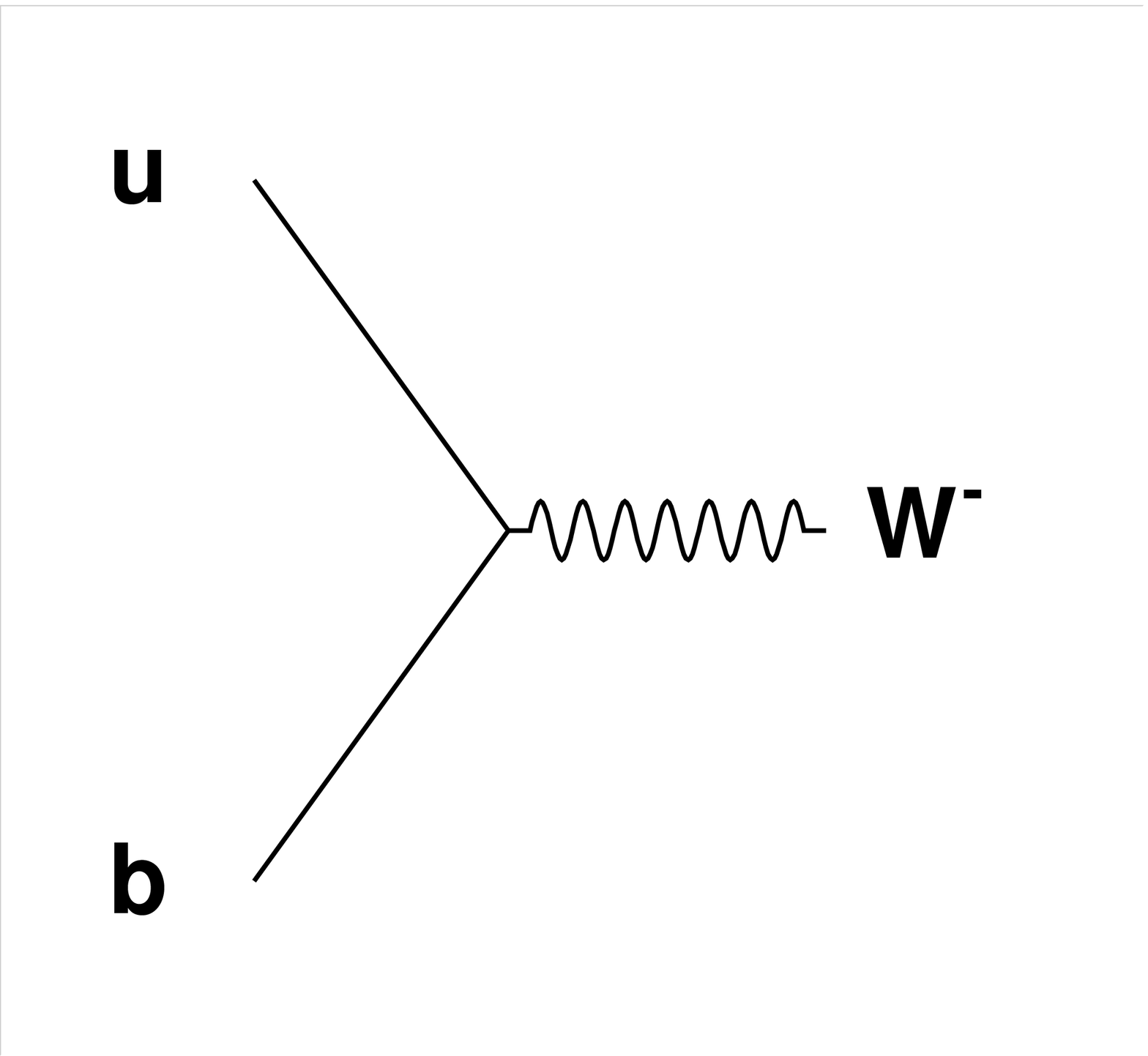,height=1.5in}
\caption{Transition diagrams}
\label{feyn:diag2}
\end{figure}

Then the $\tau_{+}\notW$ term in the Dirac equation leads to the
transitions shown in Fig.~\ref{feyn:diag2}.
\bc
\ec
The parameters $V_{ij}$ are not known {\em a priori}, but rather must be
set by experiment.
The most precise determination of any of the $V_{ij}$ comes from a nuclear
experiment
-- namely, nuclear $\beta$-decay\cite{towner}, from which one extracts a
value for $V_{11}$.\\
\noindent
The study of the CKM matrix is an important component of electroweak
physics. Unfortunately, we will not
be able to discuss these studies in any detail in these lectures. However,
a few comments are worth
making here:
\medskip
\noindent (i) The elements of ${\hat V}$ are often labeled  by the
relevant charged currents transition:\\
\bc
$V_{11} \rightarrow V_{ud}$\\
$V_{12} \rightarrow V_{us}$\\
{\em etc.}\ \ \ ,\\
\ec
that is, the element $V_{11}$ governs the strength of the transition $d\to
u +W^-$,  {\em etc.}
\medskip
\noindent (ii)The phase in the matrix $e^{i\delta}$ is responsible for
CP-violation.  In order for such a
phase to appear in ${\hat V}$, one
needs  at least 3 generations of massive quarks.
\medskip
\noindent (iii) For $m_{\nu}$=0 as assumed by the Standard Model, there is
no CKM matrix for leptons.  The
neutrinos weak and mass eigenstates are identical, and that is enough to
evade the need for a CKM
mixing matrix. However, now that neutrino oscillations have been observed
and we know neutrinos have mass, we are
forced to write down an analog of the CKM matrix for leptons. The study of
the corresponding mixing angles and phase(s)
are now a topic of intense study in nuclear and particle
physics\footnote{If neutrinos are Majorana particles, there exist
additional CP violating phases beyond the single phase associated with
mixing of three generations of massive Dirac particles.}.
\medskip
\noindent (iv) The neutral current interactions are independent of the
$V_{ij}$, and they entail no transitions
among quark generations. In short, the Standard Model forbids flavor
changing neutral current (FCNC) at lowest order
in $g$.
\subsection{Spontaneous Symmetry Breaking, Mass Generation, and the Higgs}
So far not much has been said much about quark, lepton, and gauge boson
masses.
Naively, one might think that simply putting in mass operators into the
Dirac equation as in Eqs. (\ref{eq:diracleft2},\ref{eq:diracright2}) would
take care of fermion masses.  However, life is more complicated than that.
One needs to work with the
Lagrangian from which the Dirac equation is derived:
\begin{equation}
\label{eq:fermion}
L={\bar\psi}(i\notcder-m)\psi\ \ \ .
\end{equation}
It is straightforward to show that the mass term has the following
decomposition:
\begin{equation}
\label{eq:fermionmass}
m{\bar\psi}\psi=m{\bar\psi_{L}}\psi_{R}+m{\bar\psi_{R}}\psi_{L}\ \ \ .
\end{equation}
Now, under SU(2)$_{L}$ we have:
\bea
\psi_{L}& \rightarrow& e^{i\overrightarrow{\alpha} \cdot {\tau \over 2}}
\psi_{L}\\
\psi_{R} &\rightarrow& \psi_{L}\ \ \ ,
\eea
so that the mass term breaks SU(2)$_{L}$ invariance in the Lagrangian.  A
similar problem arises for the
gauge bosons.  For a massless gauge boson like the $\gamma$, the Lagrangian
is given by:
\begin{equation}
L_{\gamma}=-{1 \over 4} F_{\mu\nu}F^{\mu\nu}\ \ \ ,
\end{equation}
where
\begin{equation}
F_{\mu\nu} = [D_{\mu},D_{\nu}] =
\partial_{\mu}A_{\nu}-\partial_{\nu}A_{\mu}\ \ \ .
\end{equation}
Since $F_{\mu\nu}$ is built from $D_{\mu}$'s, it is manifestly gauge
invariant.  Similarly, one could
write down an SU(2)$_{L}$ gauge-invariant Lagrangian for the weak gauge
bosons in the limit that they
were massless:
\begin{equation}
\label{eq:gaugekin}
L_{GB}=-{1 \over 4}\sum_{\alpha=1}^{3} F^{a}_{\mu\nu}F^{a\mu\nu}\ \ \ ,
\end{equation}
where
\begin{equation}
F_{\mu\nu}^{a} =
\partial_{\mu}W_{\nu}^{a}-\partial_{\nu}W_{\mu}^{a}+g\epsilon^{abc}W^{b}_{\mu}W^
{c}_{\nu}
\end{equation}
The extra term in $F_{\mu\nu}^{a}$ is needed to maintain invariance under
the non-Abelian
transformation (\ref{eq:su2b}).  To get the masses for the $W^{\pm}$ and
$Z^{0}$, one would naively think to
add the mass term:
\begin{equation}
\label{eq:gaugemass}
L_{M}={1 \over 2}M^{2}W^{a}_{\mu}W^{\mu a}
\end{equation}
to (\ref{eq:gaugekin}).  However (\ref{eq:gaugemass}) is again not
invariant under (\ref{eq:su2b}).  Since our goal
is to write a Lagrangian having SU(2)$_{L}
\times$U(1)$_{Y}$ gauge symmetry, adding mass terms as in
(\ref{eq:fermionmass}) and
(\ref{eq:gaugemass}) would be a disaster. How then, do we give particles
their masses?
The resolution
is the so-called Higgs Mechanism\cite{higgs}.  The idea is introduce a new
particle described by a two component
vector which transforms as a doublet under SU(2)$_{L}$:
\bea
\Phi&=&\begin{pmatrix} \phi^{+} \cr \phi^{0} \end{pmatrix}\\
Y(\Phi)&=& 2(Q-T_{3}^L)=1\ \ \ .
\eea
Using $\Phi$, we add new terms to the gauge boson (GB)-fermion(F)
SU(2)$_{L} \times$ U(1)$_{Y}$
invariant Lagrangian:
\begin{equation}
L=L_{GB}+L_{f}+L_{H}+L_{Y}\ \ \ ,
\end{equation}
where $L_{GB}$ is given by (\ref{eq:gaugekin}) for both the $W^{a}_{\mu}$
and $B_{\mu}$ and $ L_{f}$
is given by (\ref{eq:fermion}) with the mass term removed.  The new terms
are:
\bea
\label{eq:higgs}
L_{H}&=&(D_{\mu}\Phi)^{\dagger}(D_{\mu}\Phi)-V(\Phi)\\
V(\Phi)&=& -\mu^{2}\Phi^{\dagger}\Phi + \lambda(\Phi^{\dagger}\Phi)^{2}\\
D_{\mu}\Phi&=&(\partial_{\mu} -
i g {{\vec\tau} \over 2} \cdot {\vec W_{\mu}} - i g^{\prime} {Y \over 2}
B_{\mu})\Phi\ \ \ ,
\eea
and
\begin{equation}
L_{Y} = f^{(e)}{\bar\Psi^{L}_{l}}\Phi  \psi^{R}_{e} + f^{(u)}{\bar\Psi^{L}_{q}}
\widetilde{\Phi}  \psi^{R}_{u} +  f^{(d)}{\bar\Psi^{L}_{q}}\widetilde{\Phi}
\psi^{R}_{d}\ \ \ ,
\end{equation}
where
\begin{equation}
\widetilde{\Phi}=i\tau_{2}\Phi^{*}\ \ \  {\mbox{and}} \ \ \
Y(\widetilde{\Phi})= -1\ \ \ .
\end{equation}
\noindent There are a few observations to make:
\medskip
\noindent (1) The terms in $L_{H}$ and $L_{f}$ are all invariant under
SU(2)$_{L} \times$U(1)$_{Y}$
transformations.\\
\noindent (2) The first term in (\ref{eq:higgs}) involves interactions of
the type:
\begin{equation}
\label{eq:higgsgauge}
g^{2}\Phi^{\dagger} {\overrightarrow{\tau} \over 2} \cdot \overrightarrow{W_{\mu}}\Phi + g^{\prime 2}\Phi^{\dagger} {Y \over 2} B_{\mu}{Y \over 2} B^{\mu} \Phi
\end{equation}
plus cross terms involving ${\vec\tau}\cdot{\vec W}_\mu$ and $B_\mu$.\\
\noindent (3) The potential $ V(\Phi)$ has a minimum for $\Phi \not{=}0$ if
$\mu^{2}$, $\lambda $  $> 0$.
Specifically, the minimum occurs for $\Phi^{\dagger}\Phi = \mu^{2}(2\lambda)$.
The field $\Phi$ likes to sit at this point.  It is energetically favorable
to do so.  That means the
expectation value of $\Phi$ in the ground state of the universe, that is,
the vacuum, should be
non-zero:
\begin{equation}
<0|\Phi|0> \noteq 0\ \ \ .
\end{equation}
In short, $\Phi$ has a non-zero expectation value (VEV).
One can arrange things to put $<0|\Phi|0>$ at the minimum of the potential
by taking
\begin{equation}
<0|\Phi|0> = \Phi_{0}=\begin{pmatrix} 0  \\ {v / \sqrt{2}} \end{pmatrix}
\noteq 0
\end{equation}  
and letting
\begin{equation}
\label{eq:higgsvev}
\Phi=\Phi_{0} + \delta\Phi\ \ \ ,
\end{equation}
where $\delta\Phi$ denotes fluctuations of this field, called the Higgs
field, about $\Phi_{0}$.
Now observe what happens if one substitutes (\ref{eq:higgsvev}) into
(\ref{eq:higgsgauge}) one obtains
\bea
{v^{2} \over 8} &[& g^{2} (W^{1}_{\mu}W^{\mu 1 } W^{2}_{\mu}W^{\mu 2 } )
\nonumber \\
&+&(g W^{3}_{\mu} -
g^{\prime}B_{\mu})(g W^{\mu 3 } - g^{\prime}B^{\mu}) + O(\delta\Phi)
\nonumber \\
&=& {v^{2} \over 8}(2
g^{2}W^{\dagger}_{\mu}W^{\mu - } +(g^{2} + g^{\prime 2}) Z_{\mu} Z^{\mu}) +
O(\delta\Phi) \ \ \ .
\eea
Note that these terms look suspiciously like mass terms if makes the
identifications:
\begin{equation}
{v^{2}g^{2} \over 4} = M^2_{W}\ \ \  {\mbox{and}}\ \ \  {v^{2} \over 4}(g^{2} + g^{\prime 2}) = M^{2}_{Z}
\end{equation}
\begin{equation}
{M_{W} \over M_{Z}}= {g \over \sqrt{g^{2}+ g^{\prime 2}}} = \cos \theta_{W} \ \ \ .
\end{equation}
The beauty of this idea is that (a) one gets the masses for the $W^{\pm}$
and $Z^{0}$ without spoiling
the SU(2)$_{L} \times $U(1)$_{Y}$ invariance of the Lagrangian, and (b) the
photon stays massless.  One
says that the SU(2)$_{L} \times$ U(1)$_{Y}$ symmetry is {\em spontaneously
broken} by the vacuum down to
U(1)$_{EM}$ by giving the Higgs a vacuum expectation value:
\bc
$SU(2)_{L} \times U(1)_{Y} \rightarrow U(1)_{EM}\ \ \ .$\\
\ec
Similarly one will notice from the Lagrangian $L_{Y}$ that leptons and
quarks get masses from
$\Phi_{0}$ without spoiling the SU(2)$_{L} \times$ U(1)$_{Y}$ invariance of
the fermion part of the
theory.

Now, there is one last important observation.  The gauge boson masses,
$\theta_{W}$, $e$, $g$, $g^{\prime}$ and $v$ are all inter-related.  In
fact, only {\em three} of these
parameters are independent.  Once three parameters have been specified the
others are now
determined. For example, we may choose the gauge boson masses and EM charge
as the independent
parameters. Then the other parameters are determined as follows:
\bea
(M_{W},M_{Z},e) \rightarrow  \cos \theta_{W} & = & {M_{W} \over M_{Z}} \nonumber\\
g & = & {e \over \sin \theta_{W}}\\
v^{2} & = &{M_{W}^{2}(M_{Z}^{2} - M_{W}^{2}) \over \pi \alpha M_{Z}^{2}} \nonumber \\
& \approx & (242\  {\rm GeV})^{2} \nonumber\\
\eea
The quantity $v \approx 242$ GeV is known as the {\em weak scale}, that is
the scale or dimensionful
parameter associated with the symmetry breakdown SU(2)$_{L} \times
$U(1)$_{Y} \rightarrow $U(1)$_{EM}$ and the quantity which
sets the scale of the $W^{\pm}$ and $Z^{0}$ masses.
In practice, one takes the independent inputs
to be
$\alpha$, $M_{Z}$, and $G_{F}$ (the Fermi constant) measured in $\mu$-decay
which
can be related to $g$ in
terms of $\alpha$ and $M_{Z}$.\\

\subsection{Additional Observations}
Let's close this section with two observations. First, the electroweak
sector of the Standard Model
contains a sizeable number of {\em a priori} unknown parameters. They are:
\noindent (i) Gauge sector: $g$, $M_Z$, $e$  (3 parameters)
\medskip
\noindent (ii) Higgs sector: $M_H$ 	(1)
\medskip
\noindent (iii) Fermion sector: lepton and quark masses (9) and CKM angles
and phase (4)
\medskip
Hence, the electroweak sector presents 17 independent parameters which must
be taken from experiment.
This is a fairly unsatisfying situation. In fact, one motivation for
seeking a larger theory in which to
embed the SM is to try and understand the origin of these parameters.

Second, you may wonder why we refer to the weak interaction as \lq\lq
weak", since its coupling
constant $g$ is not too different from the coupling for electromagnetic
interactions. The reason for
this terminology has to do with the low-energy properties of probability
amplitudes for various
processes. To illustrate, let's compare electron-muon scattering with muon
decay ($\mu^-\to\nu_\mu+
e^-+{\bar\nu}_e$). The amplitude for the former, which is purely
electromagetic, is governed by the
photon propagator, which goes as $1/q^2$, where $q_\mu$ is the momentum
transfer in the scattering.
The $\mu$-decay amplitude, in contrast, is governed by the $W$-boson
propagator, which goes as
$1/(q^2-M_W^2)$. Since the energy released in $\mu$-decay is tiny compared
to $M_W$, this amplitude
goes as $1/M_W^2$. Thus, the ratio of the two amplitudes is
\begin{equation}
{{\mbox{WEAK}}\over{\mbox{EM}}}\sim {q^2\over M_W^2} << 1
\end{equation}
at low-energies. In short, low-energy weak interactions are \lq\lq weak" in
comparison to EM
interactions because the
$W^\pm$ is quite massive while the $\gamma$ is massless. Note, however,
that at higher energies, the
strengths of the two interactions may become comparable.

\section{Low-Energy Tests of the Standard Model}
So far, considerable attention has been spent on the elegant structure of
the Standard Model.
It has been demonstrated how it provides a unified framework for weak and
EM interactions based on
gauge symmetry, allows masses to be generated by spontaneous symmetry
breaking, accounts for parity and
CP-violation, and explains the disparate low-energy strengths and ranges of
the weak and EM interactions.
The Standard Model also makes a number of predictions:
	\begin{enumerate}
	\item Charged current interactions are purely left-handed
	\item Electroweak coupling strengths are universal
	\item The charged vector currents have a simple relation to isovector
	      electromagnetic currents
	\item The weak neutral current is a mixture of SU(2)$_{L}$ and
U(1)$_{EM}$ currents,
with the degree of
mixing characterized by $\sin^{2}\theta_{W}$
	\item Neutral current interactions conserve flavor
	\end{enumerate}
and so on. Low-energy experiments in nuclei and atoms have played
important role in
establishing these properties to a high degree of accuracy.  The following
is  a review of some of the
classic ways in which this has been done.
\medskip
\subsection{Muon Decay}
\begin{figure}[h]
\includegraphics[angle=0,scale=0.23]{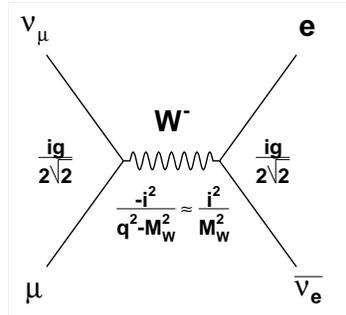}
\caption{Muon decay}
\label{mu:diag4}
\end{figure}

The decay $\mu^{-} \rightarrow \nu_{\mu} e^{-} \bar{\nu_{e}}$ is governed
at lowest orderby the amplitude in Fig.~\ref{mu:diag4}.
\bc
\ec
The amplitude is proportional to $g^2/M_W^2$, a combination of constants
that is related to the Fermi constant:
\begin{equation}
\label{eq:fermi1}
\frac{G_{F}} {\sqrt{2}} = \frac{ g^{2}}{8M_{W}^{2}}
\end{equation}

\begin{figure}[h]
\hspace{.1cm}\psfig{figure=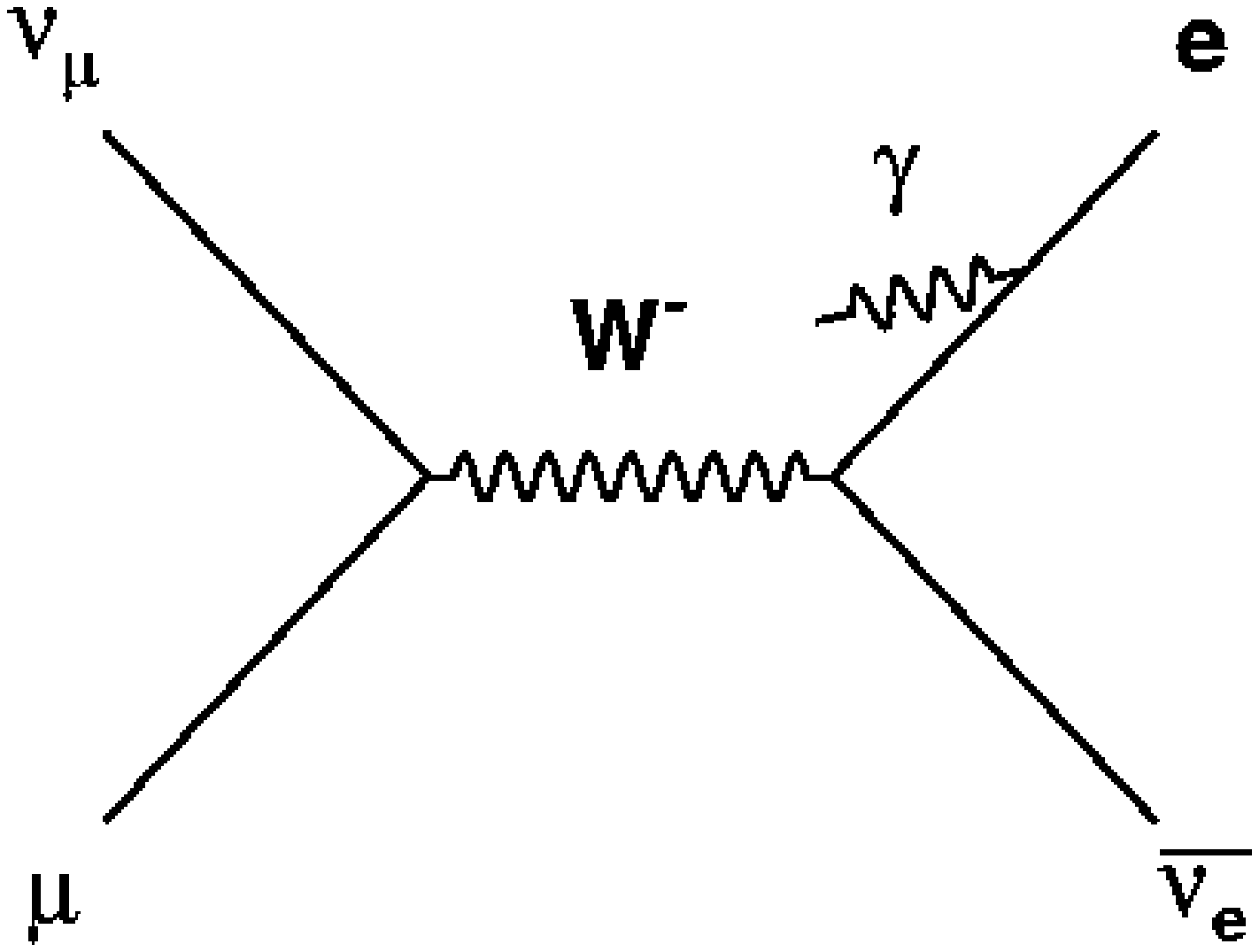,height=1.5in}
\psfig{figure=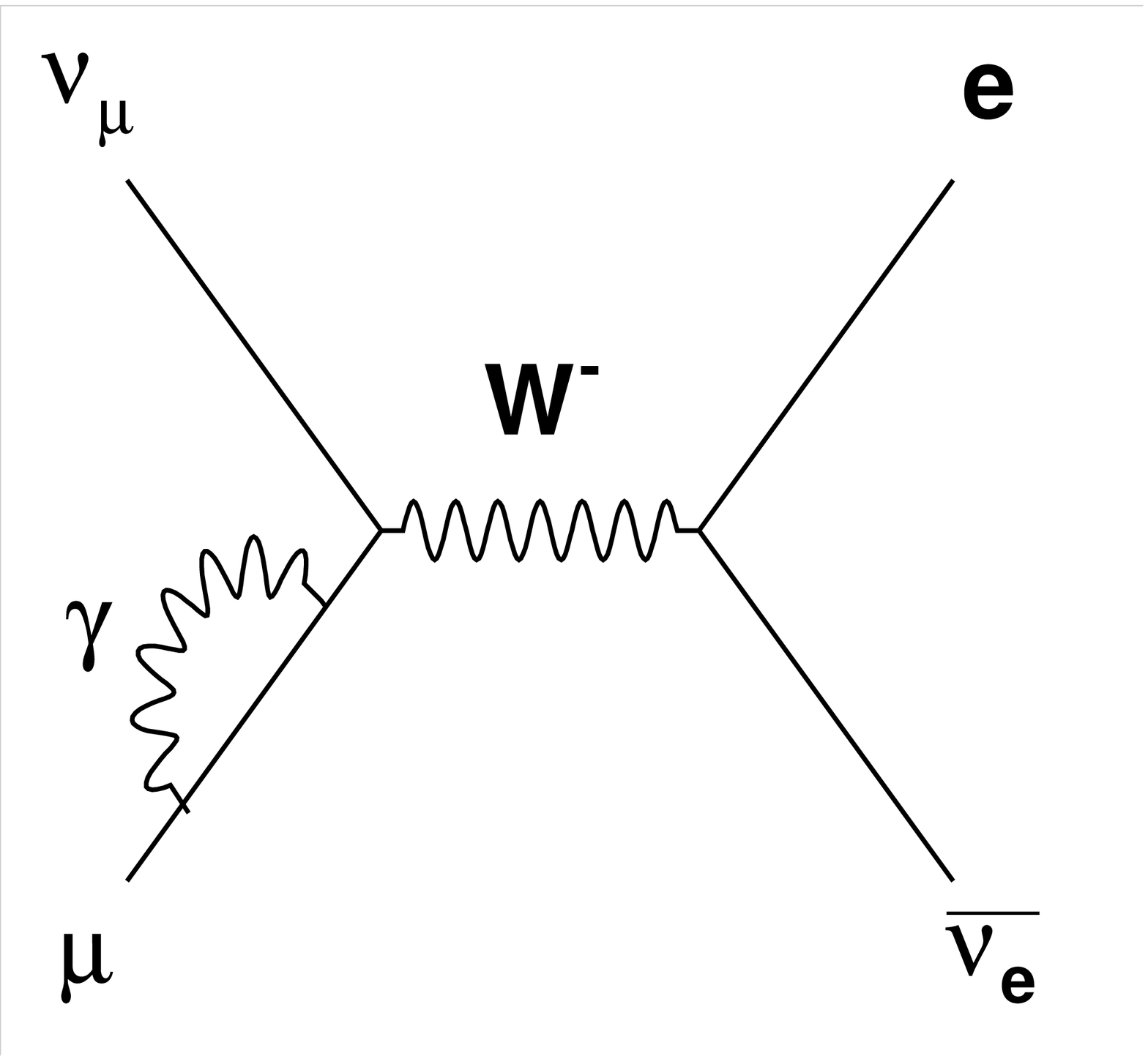,height=1.5in}
\psfig{figure=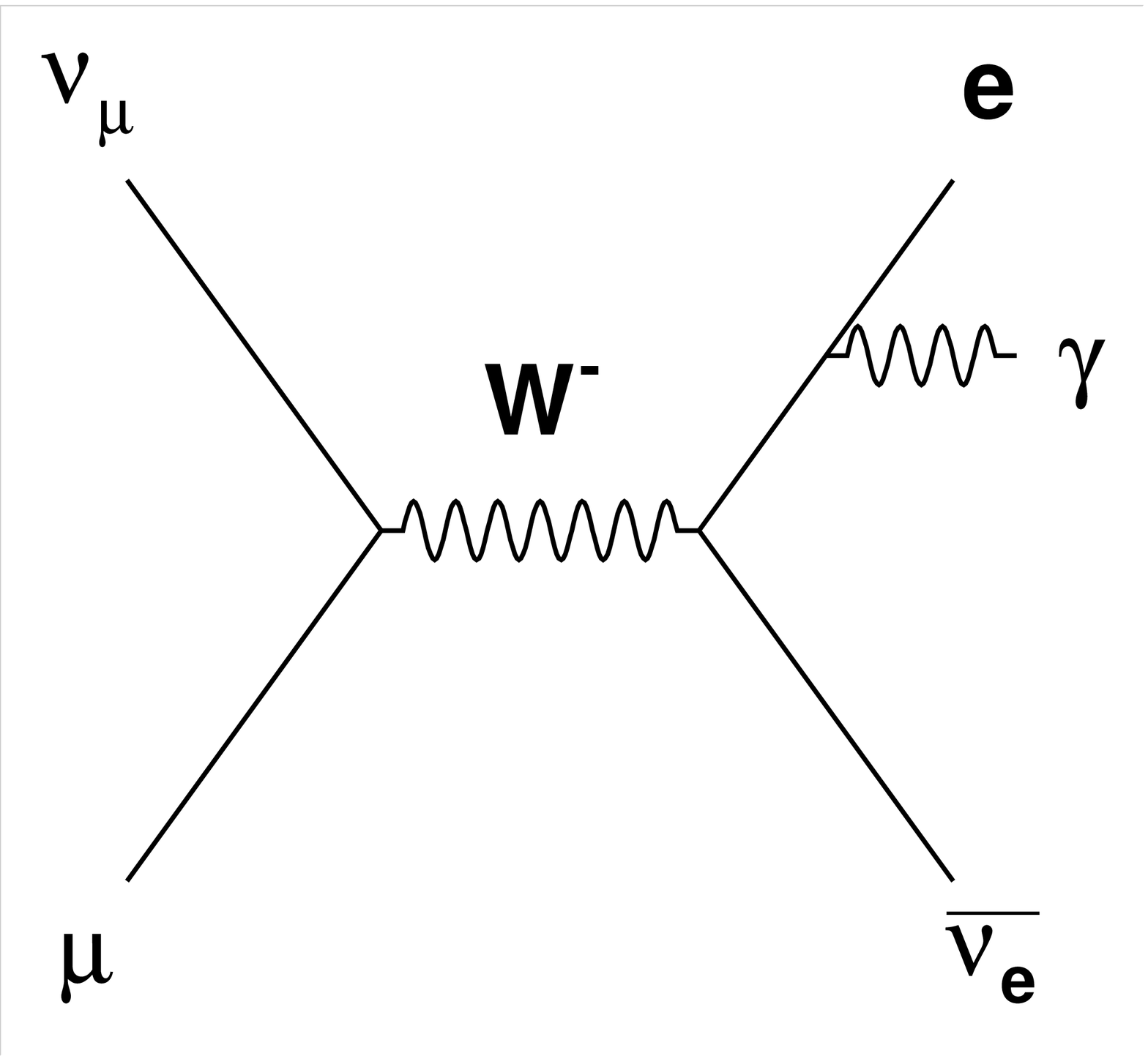,height=1.5in}
\caption{QED radiative corrections to muon decay}
\label{radcorr:diag5}
\end{figure}

Higher order corrections to this amplitude arise from $\gamma$ exchanges as
illustrated in
Fig.~\ref{radcorr:diag5}.

The effect on the total muon decay rate of these purely QED radiative
correction can be
computed precisely. The result is
\begin{equation}
\tau^{-1}_\mu = \Gamma = {G^{2}_{F} m^{5}_{\mu} \over 192 \pi^{3}}\left[ 1-{\alpha \over 2\pi}(\pi^{2}-{25 \over 4})\right] f({ m^{2}_{e} \over m^{2}_{\mu}})(1 + {3 \over 5}{m^{2}_{\mu} \over m^{2}_{W}}) 
\end{equation}
\begin{equation}
f(x)=1 -8x +8x^{3}-x^{4}+12x^{2}\ln ({1 \over x})  \nonumber
\end{equation}
The point of writing this expression is that if $\tau$ can be measured very
accurately,
$G_{F}$ can be extracted extremely precisely  since all the QED effects can
be computed.  Moreover,
since $G_{F}$ is related to $g$ and $M_{W}$, it can be used as one of the
three inputs into the gauge
sector of the Standard Model.
The muon life time is now known to the precision:
\bc
$\tau_{\mu} = 2.197035(40) \times 10^{-6}$ 
\ec
From which one obtains\cite{pdg}:
\bc
$G_{F}=1.16637(1) \times 10^{-5}\  {\rm GeV}^{-2}$
\ec
For reasons which will become apparent later, let's denote this value of
$G_{F} \rightarrow G_{\mu}$.
It is an experimental parameter.  It can be related it to the gauge sector
parameters of the Standard Model
using equation (\ref{eq:fermi1}).  However, at this level of precision, one
must take into account electroweak
radiative corrections shown in Fig.~\ref{ewcorr:diag5}:
\begin{figure}
\hspace{.1cm}\psfig{figure=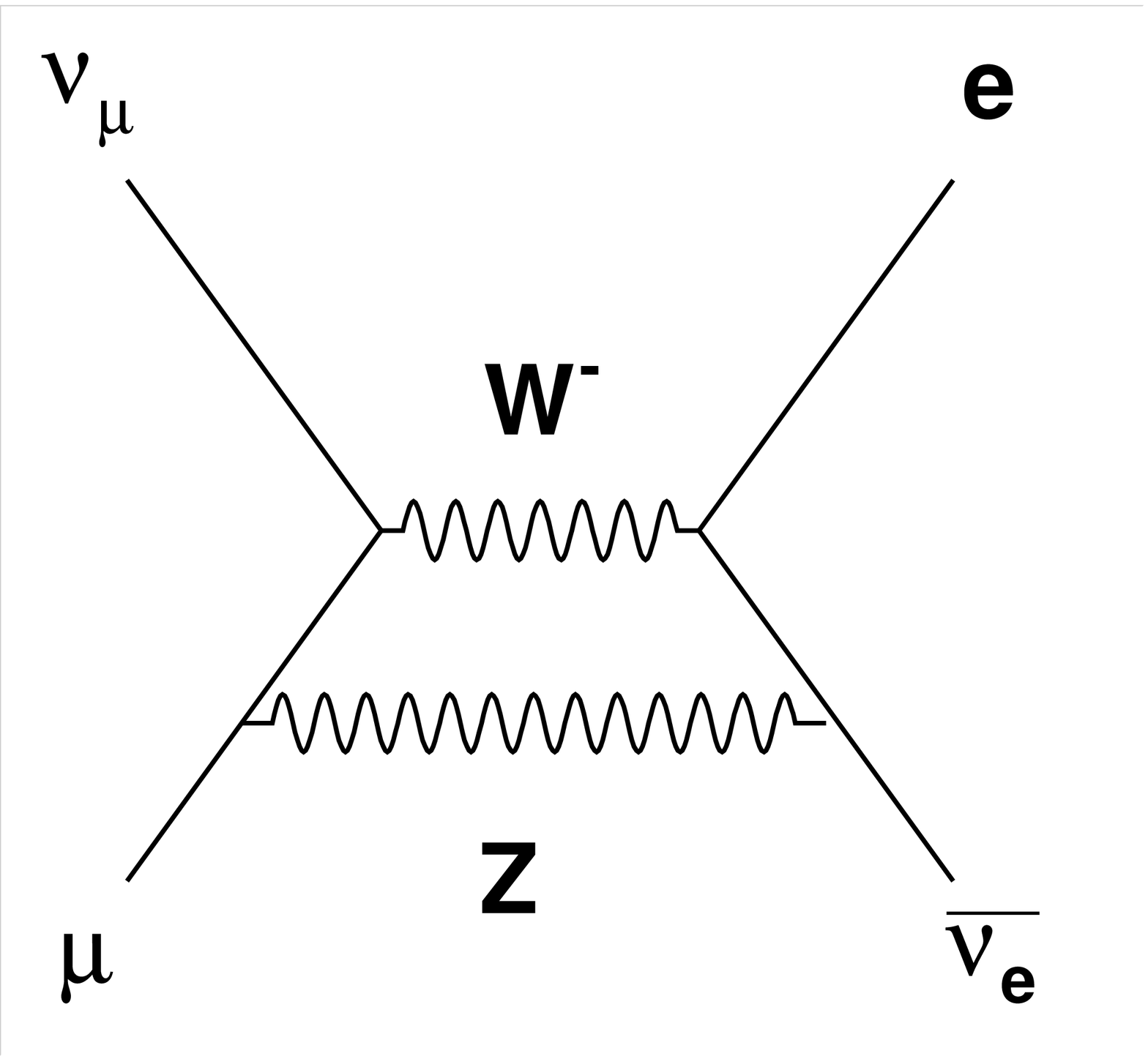,height=1.5in}\psfig{figure=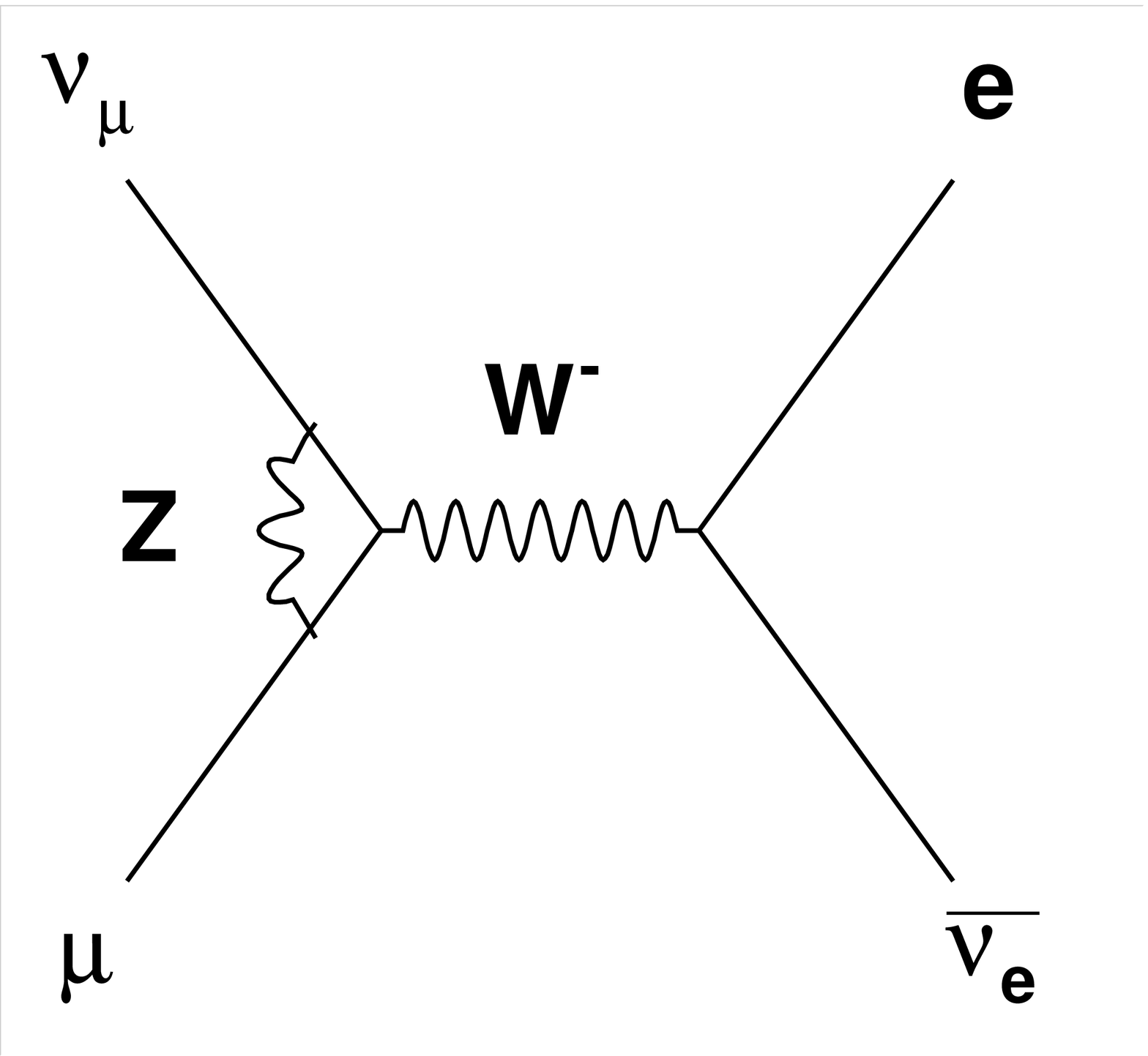,height=1.5in}
\caption{Electroweak radiative corrections to muon decay.}
\label{ewcorr:diag5}
\end{figure}
The presence of these effects, which can be computed in the Standard Model,
is to modify (\ref{eq:fermi1}):
\begin{equation}
\label{eq:fermi}
\frac{g^{2}}{8M_{W}^{2}}(1 + \Delta r_{\mu}) = \frac{ G_{\mu}}{\sqrt{2}}\ \ \ ,
\end{equation}
where $\Delta r_\mu$ denotes the corrections induced by the processes in
Fig.~\ref{ewcorr:diag5}.
One can then use relation (\ref{eq:fermi}) to compute other electroweak
observables where the value of $g$
is needed very precisely, or to test the self-consistency of the
electroweak measurements.  Letting
$g=e/\sin\theta_{W}$, $e^{2}=4\pi\alpha$, and treating $\Delta r_\mu$ as
small leads to
\begin{equation}
G_{\mu}= {\pi \alpha \over \sqrt{2} M_{W}^{2} \sin^{2} \theta_{W}(1-\Delta
r_{\mu})}
\end{equation}
One can then take $M_{W}$,$\alpha$, and $\sin^{2} \theta_{W}$ from experiment,
compute $\Delta r_{\mu}$, and ``predict'' $G_{\mu}$.  In this case, one
obtains\cite{marciano1}:
\bc
$G^{SM}_{\mu} = 1.1661(\mp 0.0018)\begin{pmatrix}+0.0005 \\ -0.0004
\end{pmatrix} \times 10^{-5} {\rm GeV}^{-2}\ \ \ .$
\ec
The level at agreement between this prediction and the result of
$\tau_{\mu}$ is impressive, and
suggests that the way one puts together the gauge sector of the Standard
Model is right.\\
Other features of the Standard Model can be tested with $\mu$-decay.  For
example, the outgoing
$e^{-}$ in $\mu^{-}$-decay should be left-handed.  In the limit where one
neglects $m_{e}$, this
implies $h(e^{-})=-1$.  Similarly, the Standard Model predicts
$h(e^{+})=+1$ in $\mu^{+}$-decay.  The
experimental limits are\cite{commins}:
\bc
$h(e^{-})=-0.89 \pm 0.28$\\
$h(e^{+})= +.94 \pm 0.08$\\
\ec
consistent with the Standard Model.\\
Another handle on the Standard Model is to consider the decay of polarized
muons.
The rate can be written in terms of the so-called Michel
parameters\cite{michel}.
The rate is a function of:
\begin{equation}
\nonumber
X= {|\vec{p_{e}}| \over |\vec{p_{e}},max|}
\end{equation}
and $\theta$, the angle between $\vec{p_{e}}$ and the $\vec{\mu}$ spin.
The parameters $\rho$ and $\delta$ characterize the spectral shape for
large $X$, and $\eta$
characterizes the low energy region.  The parameter $\xi$ is an asymmetry
parameter
that characterizes the strength of a term in the rate proportional to
$\cos\theta$.
The Standard Model predictions (for purely left-handed weak interactions)
and experimental results are\cite{commins}:
\begin{center}
\begin{tabular}{|c|c|c|}\hline
Parameter	& Experiment		& Standard Model\\ \hline
$\xi$		& -1.0045 $\pm$ 0.0086	& -1.0	 \\ \hline
$\rho$		& 0.7578 $\pm$	0.0026	& $\frac{3}{4}$     \\ \hline
$\eta$		& -0.007 $\pm$0.013	& 0	   \\ \hline
$\delta$ 		& 0.7486 $\pm$0.0038	& $\frac{3}{4}$    \\ \hline
\end{tabular}
\end{center}
Again, the level of agreement with the Standard Model is quite impressive.
As in the case of $G_{\mu}$, one can use experimental results for the
Michel parameters to put limits
on possible physics ``beyond'' the Standard Model.  Plans are currently
being made to measure  $G_{\mu}$
even more precisely, and a new experiment to determine the Michel
parameters is being performed at TRIUMF
by the TWIST collaboration.
\subsection{Pion Decay and Lepton Universality}
The reactions:
\bc
$\pi^{+} \rightarrow e^{+} \nu_{e}$\\
$\pi^{+} \rightarrow \mu^{+} \nu_{\mu}$\\
\ec
occur at  lowest order  when a $u\bar{d}$ quark pair annihilates into  a
$W^{+}$.
According to the Standard Model, the coupling of the $W^{+}$ to a
$u\bar{d}$ is proportional to:
\begin{equation}
{g \over 2 \sqrt{2}} V_{ud} \bar{d} \gamma_{\mu}(1-\gamma_{5})u = J^{CC}_{\mu}
\end{equation}
where \lq\lq CC" stands for \lq\lq charged current."
The matrix element of $J^{CC}_{\mu}$ in the decay of the $\pi^{+}$ is:
\begin{equation}
<0|J^{CC}_{\mu}|\pi^{+}(q)> = i {  F_{\pi}  \over \sqrt{\omega_{\pi}}}q_{\mu}\ \ \ ,
\end{equation}
where $F_{\pi}\approx 93$ MeV is the $\pi$ decay constant.
It parameterizes all the strong interaction(QCD) physics responsible for
binding the $u\bar{d}$ into a
$\pi^+$ --  physics one can't calculate reliably.  Hence, $F_{\pi}$ is
taken from experiment.
The total rate for  $\pi^{+} \rightarrow l^{+} \nu_{l}$ in the Standard
Model is:
\begin{equation}
\Gamma = G_{\mu}^{2} {|V_{ud}|^{2} \over 4 \pi} F_{\pi}^{2} m^{2}_{l}
m_{\pi}\left[1- \Bigl({ m^{2}_{l} \over
 m^{2}_{\pi}}\Bigr)\right]^{2}
\end{equation}
note that in the ratio of partial rates
\begin{equation}
R_{e/\mu} = {\Gamma (\pi \rightarrow e \nu) \over \Gamma (\pi \rightarrow
\mu \nu)}
\end{equation}
the dependences on $G_{\mu}, |V_{ud}|$ and $F_{\pi}$ cancel out.  The only
difference in the
$\pi^{+}
\rightarrow e \nu$ and $\pi^{+} \rightarrow l \nu$ rates came from the
lepton masses.  Thus, the
Standard Model makes a very precise prediction for $R_{e/\mu}$, even after
including electroweak
radiative corrections of the type that enter $\Delta r_{\mu}$.\\
Importantly, the ratio is insensitive to the $\pi^{\pm}\to W^{+}$ component
of the decay matrix element,
which is the same for both types of lepton final state.  Hence, determining
$R_{e/\mu}$ is a way to
test the Standard Model prediction that the $W^{+}
\rightarrow e^{+} \nu_{e}$ and $W^{+} \rightarrow \mu^{+} \nu_{\mu}$
couplings are the same, that is,
universal.\\
Recently, $R_{e/\mu}$ has been determined very precisely at PSI\cite{mupsi}
and TRIUMF\cite{mutriumf}.  The comparison
of experiment (exp) with the Standard Model prediction (SM)
yields\cite{marciano2}:
\begin{equation}
{ R^{\mbox{exp}}_{e/\mu} \over R^{\mbox{SM}}_{e/\mu}} = 0.9958 \pm
0.0033({\mbox{exp}}) \pm
0.0004({\mbox{theory}})\ \ \ .
\end{equation}
Note the impressive precision both of the experiment($\sim$ 0.3$\%$) and
theory (0.04$\%$).
The result is a clear indication of the $e$-$\mu$ universality of the
Standard Model.
\subsection{Nuclear $\beta$-Decay: Lepton-hadron CC Universality and CKM
Mixing}
One of the classic and critically important nuclear physics processes used
to test the Standard Model
is nuclear $\beta$-dacay.  Nuclear $\beta$-decay can occur via one of the
following processes:
\bea
n &\rightarrow& p +e^{-} + \bar{\nu_{e}} \\
p &\rightarrow& n +e^{+}+\nu_{e}\ \ \
\eea
where the second reaction occurs when a proton is bound in a nucleus.
As in the case of the $\pi$-decay, the currents on the left hand side of
these diagrams are given by:
\begin{eqnarray}
{\rm Charge\ \ lowering} &{\rm :}&  {g V_{ud} \over 2 \sqrt{2}} \bar{d}
\gamma_{\mu} (1 - \gamma_{5}) u\\
{\rm Charge\ \  raising}\ &{\rm :}&   {g V_{ud} \over 2 \sqrt{2}}
\bar{u}\gamma_{\mu} (1 - \gamma_{5}) d\ \ \ .
\end{eqnarray}
For the moment, let's focus on the vector part of the quark currents, omitting the constants $g$,$V_{ud}$, {\em etc.}. One has:
\begin{equation}
\label{eq:ccraise}
J^{+}_{\mu} = \bar{u} \gamma_{\mu} d
\end{equation}
\begin{equation}
\label{eq:cclower}
J^{-}_{\mu} = \bar{d} \gamma_{\mu} u \ \ \ .
\end{equation}
Compare these currents  with the isovector EM current:
\begin{equation}
\label{eq:emcurrent}
J^{T=1}_{\mu}(EM) ={1 \over 2}( \bar{u} \gamma_{\mu} u - \bar{d}
\gamma_{\mu} d)
\end{equation}
A compact way of writing Eqs. (\ref{eq:ccraise}-\ref{eq:emcurrent}) is to
use:\\
\bc
$Q= \begin{pmatrix}u \\ d \end{pmatrix}\ \ \
\bar{Q}=\begin{pmatrix}\bar{u} \bar{d}\end{pmatrix}$\\
\ec
So that
\bea
J^{+}_{\mu} = \bar{Q} \gamma_{\mu} \tau_{+} Q\nonumber\\
\label{eq:isotriplet}
J^{-}_{\mu} = \bar{Q} \gamma_{\mu} \tau_{-} Q\\
\nonumber
J^{T=1}_{\mu} = \bar{Q} \gamma_{\mu} {\tau_{3} \over 2} Q
\eea
Now an astute observer will realize that the set of currents
(\ref{eq:isotriplet}) form an isospin
triplet, satisfying the commutation relations:
\begin{equation}
 [I_{\pm},J^{T=1,EM}_{\mu}]= \mp J^{\pm}_{\mu}
\end{equation}
\begin{equation}
 [I_{3},J^{\pm}_{\mu}]= \pm 2J^{\pm}_{\mu}
\end{equation}
where
\begin{equation}
I_{k} = \int{d^{3}x \bar{Q} \gamma_{0} {\tau_{k} \over 2} Q}
\end{equation}
In short, $ \{J_\mu^{\pm},J^{T=1,EM}_{\mu}\}$ satisfy the same commutation
relations as
$\{I^{\pm},I_{3}\}$.   Consequently, matrix elements of the currents
(\ref{eq:ccraise}-\ref{eq:emcurrent}) must be related in the same way as
matrix elements of the isospin
operators (up to angular momentum properties).  This property of the
Standard Model is known as  the
\lq\lq conserved vector current" property, or CVC.  It implies that:\\
\noindent (i) Matrix elements of $J^{\pm}_{\mu}$ between nuclear states of
different total isospin (I)
must vanish.\\
\noindent (ii) Matrix elements of $J^{\pm}_{0}$ satisfy:\\
\begin{equation}
\label{eq:matrixelement}
<I,I_{z} \pm 1 | J^{\pm}_{0} | I,I_{z}> = [(I \mp I_{z})(I \pm I_{z} +1]^{1 \over 2}
\end{equation}
\noindent at $q^2=0$.
A special set of nuclear decays sensitive to the matrix element
(\ref{eq:matrixelement}) are the
\lq\lq superallowed" Fermi decays involving transitions:\\
\bc
$(J^{\pi}=0^{+},I,I_{z}) \rightarrow (J^{\pi}=0^{+},I,I_{z} \pm 1)$\\
\ec
Since the initial and final states have the same parity and zero total
angular momentum,
the axial current cannot connect them.  Moreover, since the initial and
final nuclear
spins $J_{f}=J_{i}=0$, only the vector charge operator can connect the two
states.  Letting $H_{fi}$
denote the total transition amplitude, one has
\begin{equation}
H_{fi}= { g^2 V_{ud} \over 8 M_{W}^{2}} [(I \mp I_{z})(I \pm I_{z}+1)]^{1
\over 2}
\end{equation}
so that
\begin{equation}
\label{eq:hfi}
|H_{fi}|^{2}= {G^{2}_{\mu} |V_{ud}|^{2} \over 2} [(I \mp I_{z})(I \pm I_{z}+1)]
\end{equation}
at $q^{2}=0$.  Amazingly, the value of $|H_{fi}|^{2}$
for {\em any} superallowed decay depends only on the muon decay Fermi
constant, $|V_{ud}|^{2}$ from the
CKM matrix, and the isospin factor.  In the expression for (\ref{eq:hfi}),
there appears no dependence
on the nuclear wavefunction--no matter how complex it is!  Thus, if one
takes the rate
for any superallowed decay and divides out the kinematical and the isospin
factors, one should get the same answer as for any other superallowed
decay.  If this works, then the
CVC prediction of the Standard Model is right.  Moreover, by comparing this
common rate with $G_{\mu}$,
one can test charged current universality for the leptons and quarks (the
overall strength is $\approx
G_{\mu}$) and extract the quark mixing parameter $|V_{ud}|$.  Let's see how
this works in detail:\\
The differential decay rate is, from Fermi's Golden Rule:
\begin{equation}
d\Gamma = \frac{2\pi}{ \hbar } |H_{fi}|^{2}\ \rho_{f}\ \delta(E_{0} -E_{e}-E_{\nu})
\end{equation}
where $E_{0}=$ is the energy released to the leptons and
\begin{equation}
\rho_{f}  { d^{3}p_{e} \over (2 \pi \hbar)^3} { d^{3}p_{\nu} \over (2 \pi
\hbar)^{3}}
\end{equation}
is the density of final states.
As an aside, consider for the moment the possibility\footnote{In the SM, one
has $m_\nu=0$. However, the results from a variety of neutrino oscillation
experiments have taught us
that neutrinos have mass.} that $m_{\nu_{e}} \not= 0$.  Putting in the
factors of c, we have
\begin{eqnarray}
E_{\nu}&=&(c^{2} p^{2}_{\nu}+m_{\nu}^{2} c^{4})^{1 \over 2}\\
cp_{\nu} &=& (E_{\nu}^{2} - m_{\nu}^{2}c^{4})^{1 \over 2}\ \ \ .
\end{eqnarray}
Integrating over $d^{3}p_{\nu}$ gives
\begin{equation}
d \Gamma = {2 \pi \over \hbar} |H_{fi}|^{2} {4 \pi \over c^{3}}
(E-E_{0})^{2} \left[1- {(m_{\nu} c^{2})^{2} \over (E-E_{0})^{2}}\right]^{1 \over 2} {p^2_{e} dp_{e} d\Omega \over (2 \pi \hbar)^{6}} \ \ \ .
\end{equation}
Note that for a given detector setting which accepts all counts in a solid
angle $\Delta\Omega=
\sin\theta \Delta \theta \Delta \pi$, the number of electrons counted in a
momentum slice
$\Delta p_e$ is
\begin{equation}
\Delta N \equiv N(p_{e}) \Delta p_{e} \propto (E_{0}-E_{e})^{2}\left[1 -
{(m_{\nu}c^{2})^{2} \over
(E_{0}-E_{e})^{2}}\right]^{1 \over 2} p_{e}^2\ \ \ .
\end{equation}
Now let's think about a plot of
\begin{equation}
{\sqrt{N(p_{e})} \over p_{e}} \propto (E_{o}-E_{e})\left[ 1 - {(m_{\nu}c^{2})^{2} \over (E_{o}-E_{e})^{2}}\right]^{1 \over 4}\ \ \ .
\end{equation}
This graph is known as a {\em Kurie plot}.
The curve intercepts the $x$-axis at $E_e=E_0-m_\nu c^2$. Thus,
deviations from linearity at the \lq\lq endpoint" would indicate $m_{\nu}
\noteq 0$.  One of the most
precise upper limits on neutrino mass comes from analyzing the endpoint of
the decay\cite{mnulimit}:
\begin{equation}
^{3}{\mbox{H}} \rightarrow   ^{3}{\mbox{He}} + e^{-} + \bar{\nu_{e}}\ \ \ ,
\end{equation}
which yields $m_{\bar{\nu_{e}}} \leq 15 eV$.
So until recent observations of neutrino oscillations, nuclear
$\beta$-decay confirmed at a very high
level that $m_{\nu}=0$ as implied by the Standard Model:
\begin{equation}
 {m_{\bar{\nu_{e}}} \over m_{e}} \leq 3 \times 10 ^{-5}\ \ \ .
\end{equation}
Now, back to super allowed decays.  Converting $dp_{e}$ to $dE_{e}$ and
putting in all the $\hbar$'s and
c's, we have
\bea
{d \Gamma \over dE_{e}} &=&
\left(\frac{1}{2\pi^{3}\hbar^{4}c^{6}}\right) |H_{fi}|^{2} \nonumber\\
&\times & E_{e}
[E_{e}^{2}-(m_{e}c^{2})^2]^{\frac{1}{2}}
(E_{0}-E_{e})^{2}
\left[1-\frac{(m_{\nu}c^{2})^{2}}{(E_{0}-E_{e})^{2}}\right]^{\frac{1}{2}}
\eea
after integrating over $d\Omega_{e}$.  Letting $\epsilon={ E_{e}/m_{e}c^{2}}$:\\
\begin{equation}
\label{eq:dgamde}
{d\Gamma \over d\epsilon} = \left( {m_{e}^{5} c^{4} \over 2 \pi^{3} \hbar^{7}}\right)|H_{fi}|^{2} \epsilon (\epsilon^{2}-1)^{1 \over 2}(\epsilon_{0}-\epsilon)^{2} \left[1- {\lambda^{2} \over (\epsilon_{0} - \epsilon)^{2}}\right]^{ 1 \over 2}\ \ \ ,
\end{equation}
where $ \lambda = {m_{\nu_{e}}/ m_{e}}$.
Now, we should correct for the fact that the outgoing electron wavefunction
is distorted by the nucleus and
other atomic electrons.  One can solve for the appropriate correction
factor very precisely. Let's
denote this factor by
$F(Z,\epsilon)$.  Multiplying (\ref{eq:dgamde}) by this factor and
integrating over $\epsilon$ gives the total rate:
\begin{equation}
\label{eq:super2}
\Gamma = \left( {m_{e}^{5} c^{4} \over 2 \pi^{3} \hbar^{7} }\right) |H_{fi}|^{2} f(Z)
\end{equation}
where where $f(Z)$ results from including $f(Z,\epsilon)$ in the integral.
For $0^{+} \rightarrow 0^{+}$ transitions, the Standard Model value for
$|H_{fi}|^{2}$ is
$\epsilon$-independent and is given by:
\begin{equation}
\label{eq:hfi2}
|H_{fi}|^{2} = { G_{\mu}^{2} \over 2} |V_{ud}|^{2}[(I \mp I_{z})(I \pm I_{z} + 1)]\times 2
\end{equation}
as in Eq.~(\ref{eq:hfi}), so that\\
\begin{equation}
\Gamma = ( {m_{e}^{5} c^{4} \over 2 \pi^{3} \hbar^{7} }) G_{\mu}^{2} |V_{ud}|^{2} [(I \mp I_{z})(I \pm I_{z} + 1)]f(Z)\ \ \ .
\end{equation}
The additional factor of 2 in Eq.~(\ref{eq:hfi2}) results from the purely
leptonic part of the
matrix element.
Now an aside on units. The quantity $G_{\mu}({m_{e}c^{2})^{2} /( \hbar c)^{3}}$ is dimensionless.
Thus, the dimensions of (\ref{eq:super2}) are
\begin{equation}
[\Gamma]\ \  : \ \ \left[ G_{\mu}{(m_{e}c^{2})^{2} \over ( \hbar
c)^{3}}\right]^{2} {m_{e}c^{2} \over \hbar}
\end{equation}
while
\begin{equation}
{m_{e}c^{2} \over \hbar} = {m_{e}c^{2} \over \hbar c } \times c\ \ \ .
\end{equation}
Thus, $[{m_{e}c^{2} / \hbar}] = ({\mbox{MeV}} / {\mbox{MeV}} f) \times (f / s) = {1 / s}$, so
the dimensions work out just right, with $\Gamma$ having the dimensions of
a rate.
The time dependence of decay is described by:\\
\begin{equation}
N=N_{0}e^{-\Gamma t}\ \ \ .
\end{equation}
For $t=t_{1 \over 2}$ denoting the half-life, at which time $N={N_{0}/ 2}$
we have
\begin{equation}
 \Gamma t_{1 \over 2} = \ln 2
\end{equation}
or
\begin{equation}
\label{eq:ftvalue}
ft_{1 \over 2} = \left( { 2 \pi^{3} \ln 2 \hbar^{7} \over m_{e}^{5}
c^{4}}\right) { 1 \over G^{2}_{\mu} |V_{ud}|^{2}
} {1 \over [( I \mp I_{z}) ( I \pm I_{z} +1)]}\ \ \ .
\end{equation}
The quantity in Eq. (\ref{eq:ftvalue}) is called the \lq\lq ft" value for
the decay. It turns out that for
all the experimentally studied superallowed transitions, the isospin factor
in the denominator of Eq. (\ref{eq:ftvalue})
is the same. Thus, the ft values
should be the same for all $0^{+} \rightarrow 0^{+}$ transition if CVC is
right.
In fact, nine superallowed decays have been studied. The \lq\lq ft" values
for the decays agree at an
impressive level of precision.  Thus, to an extremely high precision, the
SU(2) character of the weak charged currents are confirmed by nuclear
$\beta$-decay.
One last important feature:  the superallowed decays can test lepton-quark
universality
of the charged current weak interaction {\em and} the unitarity of the CKM
matrix.  To see how, let:
\begin{equation}
G^{\beta}_{F} = G_{\mu} |V_{ud}| (1+\Delta r_{\beta}-\Delta r_\mu)\ \ \ ,
\end{equation}
where $\Delta r_\beta$ and $\Delta r_\mu$ denote {\em radiative
corrections} to $\beta$-decay
and $\mu$-decay, respectively. The correction $\Delta r_\beta$ appears
because of higher-order
effects in the semileptonic ($d\to ue^-{\bar\nu}_e$) amplitude. On the
other hand, $-\Delta r_\mu$ appears
because the Fermi constant has been taken from the muon lifetime, and we
need to subtract out radiative corrections
to the muon decay amplitude because they are contained in $G_\mu$ but
don't affect $\beta$-decay. These corrections must be computed from the
Standard
Model.  Using the result of these calculations and the experimental muon
lifetime and $\beta$-decay
ft values, we have:
\begin{equation}
G^{\beta}_{F}=1.16637(1) \times 10^{-5} \ {\rm GeV}^{-2}
\end{equation}
and
\begin{equation}
|V_{ud}|^2= {G^{\beta}_{F} \over G_{\mu}(1+\Delta r_{\beta}-\Delta r_\mu)}=0.9740 \pm 0.0005\ \ \ .
\end{equation}
From $K_{\ell_{3}}$ and hyperon decays we have $|V_{us}|=0.2196 \pm 0.0023$
while $B$-meson decays give
$|V_{ub}|=0.0032 \pm 0.009$ .
Now unitarity of the CKM matrix requires:
\begin{equation}
|V_{ud}|^{2} + |V_{us}|^{2} + |V_{ub}|^{2} =1 \ \ \ .
\end{equation}
The experimental results give:
\begin{equation}
|V_{ud}|^{2}_{\rm exp} + |V_{us}|^{2}_{\rm exp} + |V_{ub}|^{2}_{\rm exp} =0.9968 \pm 0.0015 \ \ \ ,
\end{equation}
corresponding to a 2.2 $\sigma$ deviation from the SM requirement of CKM
unitarity.\\
Note that at the one percent level, charged current lepton-quark
universality and CKM unitarity are confirmed by
$\beta$-decay, $\mu$-decay, $K_{l_{3}}$-decays, and B-decays.
However, at the level of precision now achieved by experiment, these
features of the Standard Model
almost hang together-though not quite.  There is a hint that maybe there is
more to the electroweak
interactions than the Standard Model, and that this ``new physics'' may be
responsible for the very
tiny derivation from lepton-quark universality and CKM unitarity.
As an aside, one might wonder why one chooses to focus on nuclear decays
rather
than the decay of the free neutron.  In fact, $\tau_{n}$ has been measured
very precisely:
\begin{equation}
\tau_{n}=886.7 \pm 1.9 s\ \ \ .
\end{equation}
However, it is a more complicated matter
to extract $G^{\beta}_{F}$ from these decays.  Because the neutron has spin
$1/2$, both the vector
and axial vector weak quark currents contribute to the decay rate. The
axial vector current is not
protected by a CVC type symmetry, and it gets important renormalizations
due to the strong interaction.
At present, we cannot compute these strong interaction effects with the
kind of precision we'd need in order
to extract $|V_{ud}|^2$ from $\tau_n$ alone. In order to circumvent this
problem, one can perform measurements of
parity-violating asymmetries associated with, {\em e.g.}, the direction of
the outgoing $e^-$ relative to the
direction of neutron spin. Knowing both $\tau_n$ and one of these
asymmetries allows one to determine separately
the vector and axial vector contributions. From the former, we can obtain a
value for $|V_{ud}|^2$ without having
to worry about incalculable strong interaction effects.\\
It is only recently that experiments have begun to determine these
asymmetries with the kind of
precision needed to determine $|V_{ud}|^2$ with the same precision as
obtained from $0^{+} \rightarrow
0^{+}$ decays. At present, there is an active research program underway at
Los Alamos that will use polarized,
ultracold neutrons to measure the $\beta$-decay asymmetries. The goal of
this program is to match or even
exceed the precision on $|V_{ud}|^2$ obtained from the superallowed decays.

\subsection{Parity violating DIS and Weak Neutral currents}
So far, we've seen how low energy experiments have provided important
confirmation of several features
of the charged current weak  interaction.  What about the weak neutral current?

There exists a basic difficulty in this case. Nature has given us two neutral
currents, the electromagnetic (EM) and weak neutral current (NC):
\begin{eqnarray}
J^{EM}_{\mu} &=& \sum_{f} Q_{f} \bar{f} \gamma_{\mu} f\\
J^{NC}_{\mu} &=& \sum_{f} \bar{f} \gamma_{\mu}(g^{f}_{V} +
g^{f}_{A}\gamma_{5})f\ \ \ ,
\end{eqnarray}
where the sum runs over all species of fermions, $Q_f$ denotes the EM
charge of fermion $f$, and
$g_V^f$ ($g_A^f$) is the vector (axial vector) coupling of $f$ to the $Z^0$
boson.
In a low energy charge neutral process, amplitudes associated with both
enter.
So the problem will be how to separate them.

\begin{figure}[h]
\includegraphics[angle=0,scale=0.3]{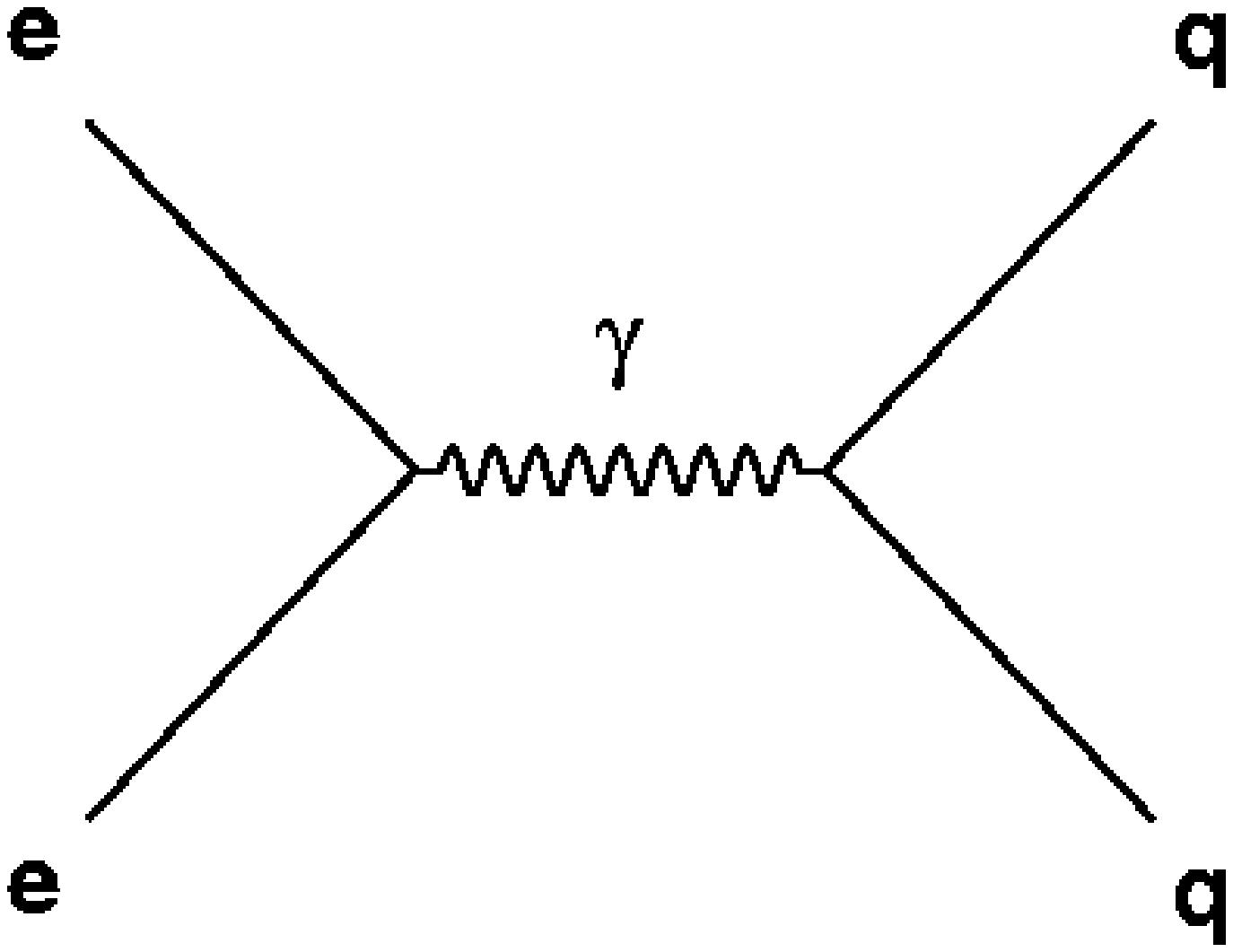}
\includegraphics[angle=0,scale=0.3]{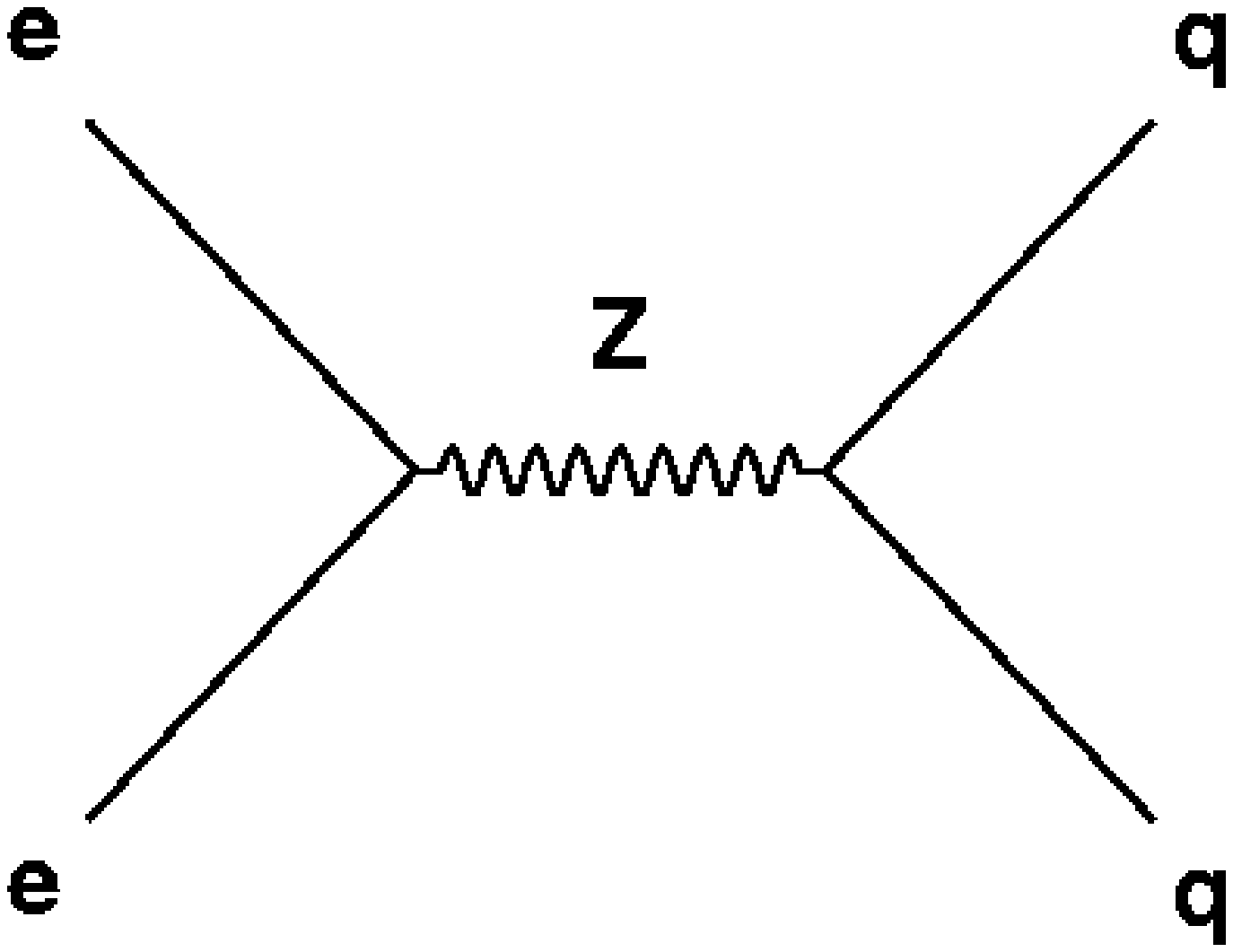}
\caption{Electron-quark scattering diagrams}
\label{eq:diag7}
\end{figure}

To be concrete, consider the scattering of electrons from quarks inside a
hadronic target.
The total amplitude for the process of $eq$ scattering is:
\begin{equation}
M = M_{EM} + M_{NC}\ \ \ ,
\end{equation}
while the cross section is $\propto |M^{2}|$:
\begin{equation}
\label{eq:NCamp}
|M|^{2} = |M_{EM}|^{2} + 2 Re(M^{*}_{EM}M_{NC}) + |M_{NC}|^{2}\ \ \ .
\end{equation}
Consequently, the neutral current cross section can be separated into three
terms:
\begin{equation}
\sigma^{\rm tot}=\sigma^{EM}+\sigma^{\rm int}+\sigma^{NC}\ \ \ .
\end{equation}
Here, $\sigma^{EM}$, $\sigma^{\rm int}$, and $\sigma^{NC}$ denote the
purely electromagnetic contribution, the
part arising from the interference of EM and weak NC amplitudes, and a
purely weak NC contribution (corresponding
to the three terms in (\ref{eq:NCamp}).
The coupling strengths entering the  amplitudes are $e$ for the EM amplitude
and ${g M_{Z}/4 M_{W}}$ for the weak NC amplitude. With this normalization
one has :
\begin{eqnarray}
g^{f}_{V}&=&2T^{f}_{3}-4Q_{f}\sin^{2}\theta_{W}\\
g^{f}_{A}&=&-2T^{f}_{3}\ \ \ .
\end{eqnarray}
This allows us to write:
\begin{equation}
|M_{EM}|^{2} \propto  \left({e^{2} \over q^{2}}\right)^{2} = \left({4 \pi \alpha \over q^{2}}\right)^{2} \ \ \ ,
\end{equation}
\bea
2Re(M^{*}_{EM}M_{NC}) &\propto& 2\left({e^{2} \over q^{2}}\right)^{2}
\left( {g M_{Z}
\over 4 M_{W}}\right)^{2} { 1
\over M_{Z}^{2}-q^2}\\
 &\rightarrow & 2\times {4 \pi \alpha \over q^{2}} {G_{\mu} \over 2
\sqrt{2}}\nonumber
\eea
at $q^2=0$, and
\begin{equation}
|M_{NC}|^{2} \propto \left[( {g M_{Z} \over 4 M_{W}})^{2} {1 \over M_{W}^{2}-q^2}\right]^{2} \propto {G_{\mu}^2 \over 8}
\end{equation}
at $q^2=0$. Now consider the relative strengths of the corresponding cross
sections at low-energies.
The ratio of the interference cross section $\sigma^{\rm int}$
to the EM cross section $\sigma^{EM}$ goes as
\begin{equation}
{\sigma^{\rm int} \over \sigma^{EM}} = {2 Re(M^{*}_{EM} M_{NC}) \over |M_{EM}|^{2}} \propto {G_{\mu} \over \sqrt{2}} {q^{2} \over 4 \pi \alpha}
\end{equation}
Letting $|q^{2}| \propto (1 {\rm GeV})^{2} \propto m_{p}^{2}$ (units where
$\hbar = c =1$) be a typical
momentum transfer for a low-energy scattering reaction, we have
\begin{eqnarray}
\alpha &\approx & {1 \over 137}\\
G_{\mu} &\sim & {10^{-5} \over m_{p}^{2}}\ \ \ ,
\end{eqnarray}
leads to:
\begin{equation}
{\sigma^{\rm int} \over \sigma^{EM}} \approx 10^{-4}\ \ \ .
\end{equation}
The magnitude of $\sigma^{NC}/\sigma^{EM}$  is even smaller.
So how is one ever going to see weak neutral current effects at low energies?
The basic idea is to use the fact that $M_{NC}$ contains pieces that are
odd under parity,
where as $M_{EM}$ is parity even. Letting
\begin{eqnarray}
V^{f}_{\mu}&=& \bar{f} \gamma_{\mu} f \\
A^{f}_{\mu}& =& \bar{f} \gamma_{\mu} \gamma_{5} f
\end{eqnarray}
one has
\begin{eqnarray}
J^{EM}_{\mu} & =&  \sum_{f} Q_{f} V_{\mu}^{f}\\
J_{\mu}^{NC} & = & \sum_{f} g^{f}_{V} V^{f}_{\mu} + \sum_{f} g^{f}_{A} A^{f}_{\mu}
\end{eqnarray}
Now, let's look at the e-q amplitudes again, as illustrated in
Fig.~\ref{eq:diag7}.
In schematic terms,  one has
\begin{eqnarray}
\label{eq:emaplitude}
M_{EM} & \sim& Q_e Q_q V_e\cdot V_q \\
M_{NC} & \sim & {G \over 2 \sqrt{2}}[g^{e}_{V}g^{q}_{V} V_{e} \cdot V_{q} + g^{e}_{A}g^{q}_{A}A_{e}
\cdot A_{q}\nonumber  \\
\label{eq:ncamplitude}
&&+ g^{e}_{V}g^{q}_{A} V_{e} \cdot A_{q} + g^{e}_{A}g^{q}_{V} A_{e} \cdot V_{q}]
\end{eqnarray}
The last two terms in $M_{NC}$ transform as pseudoscalars; they are odd
(change sign) under
parity. All the other terms in
Eqs.~(\ref{eq:emamplitude},\ref{eq:ncamplitude}) are parity-even. Thus, to
the extent that one can experimentally isolate these terms by measuring an
observable
which is parity-odd, one has a way to \lq\lq filter out" the much larger EM
interaction and get one's
hands on the effects of the weak NC.

Now consider the following experiment. A beam of electrons with with  spin
and momentum vectors parallel (helicity
$h=+1$), elastically scatter from a nucleon.  Subsequently, the incident
electron's spin is flipped, so that
its spin and momentum vectors are anti-parallel ($h=-1$).
One may then form the helicity-difference or \lq\lq left-right" asymmetry:
\begin{equation}
A_{LR} = {N(h=+1) - N(h=-1) \over N(h=+1) + N(h=-1)} \ \ \ ,
\end{equation}
where $N(h)$ denotes the number of events for electrons having incident
helicity $h$.
Since $h=\hat{s} \cdot \hat{k} \rightarrow - \hat{s} \cdot \hat{k}$ under
parity, the asymmetry $A_{LR}$
must be proportional to the parts of $|M|^{2}$ which contain pseudoscalars.
In fact, one
has
\bea
\nonumber
A_{LR} & =& { 2Re(M^{*}_{EM}M^{PV}_{NC}) \over |M_{EM}|^{2}+\cdots} \approx { 2Re(M^{*}_{EM}M^{PV}_{NC}) \over |M_{EM}|^{2}}\\
&=& {G_{\mu} q^{2} \over 4 \sqrt{2} \pi \alpha} {g^{e}_{V} g^{q}_{A} V_{e} \cdot A_{q} + g^{e}_{A} g^{q}_{V} A_{e} \cdot V_{q} \over Q_{e} Q_{q} V_{e} \cdot V_{q}}
\eea
Here I have not performed a sum over quark spins nor integrated over
energies, angles of the
outgoing quarks, {\em etc}. The expression is essentially schematic,
illustrating the basic physics
ingredients in the asymmetry.
Now, it takes a little algebra to perform this calculation, but it is
doable.  In addition, one can
analyze $e^{-}$N or $e^{-}$A scattering {\em as if} one were scattering off
individual quarks and then
summing over different types of quarks in the nucleon.  To do this one
needs to work in a kinematic
regime where 
\begin{equation}
W^{2}=(P_{target} + q)^{2} = q^{2} +2M_{t}\nu +M^{2} >> M_{T}^{2}
\end{equation}
where $\nu=q_{0}$ is the energy transfer to the target (having mass $M_T$),
which is just the energy
loss of the electron.\\
This kind of experiment is called deep inelastic scattering (DIS).  One of
the first experiments to test
the  neutral current structure of the Standard Model was carried out at
SLAC in the 1970's using
parity-violating (PV) DIS from a deuterium target\cite{slac}.  The L-R
asymmetry for this scattering is:
\begin{equation}
A^{DIS}_{LR}(\vec{e}D)={G_{\mu} q^{2} \over 4 \sqrt{2} \pi \alpha}({9 \over
10})
\left[\tilde{a_{1}}+\tilde{a_{2}}{1-(1-y)^{2} \over 1+(1-y)^{2}}\right]
\end{equation}
where:
\begin{eqnarray}
\tilde{a_{1}}&=& {1 \over 3} g^{e}_{A}(2g^{u}_{V}-g^{d}_{V})=1 - {20 \over 9} \sin^{2} \theta_{W}\\
\tilde{a_{2}} &=& {1 \over 3} g^{e}_{V}(2g^{u}_{A}-g^{d}_{A})=1 - 4 \sin^{2} \theta_{W}
\end{eqnarray}
and
\begin{equation}
y={\nu \over E_{e}}
\end{equation}
in the lab frame.
The SLAC experiment was carried out at four energies in the range:\\
\bea
16.2 &\geq& E_{e} \geq 22.2\ \ {\rm GeV}\nonumber \\
<Q^{2}> &=& 1.6\ \ ({\rm GeV} / c)^{2} \ \ \ .\nonumber
\eea
The $A_{LR}$ was measured as a function of $y$, which allowed a separation
of $\tilde{a_{1}}$ and
$\tilde{a_{2}}$.  The best fit to the results can be used to extract a
value of $\sin^{2}\theta_{W}$:
\begin{equation}
\sin^{2}\theta_{W}= 0.224 \pm 0.020 \ \ \ .
\end{equation}
Note that $\sin^{2}\theta_{W} \approx {1 \over 4}$ which implies that
$\tilde{a_{2}} \approx 0 $.
This result corresponds to $A_{LR}$ having almost no dependence on the
kinematic variable $y$.
Thus, the SLAC data is consistent with the Standard Model picture of
SU(2)$_{L}$ and U(1)$_{Y}$ mixing
in the neutral current sector with the mixing
parameter,$\sin^{2}\theta_{W}$, being close to ${1 \over 4}$.
At the time the experiment was performed, there existed competing
electroweak models which
predicted a more dramatic $y$-dependence of $A_{LR}$. These models were
ruled out by the results of
this experiment.

The SLAC value for $\sin^2\theta_W$ was later confirmed in a series of
purely leptonic experiments\cite{leptonweak}:
\bea
\nu_\mu + e &\rightarrow&  \nu_\mu + e\nonumber \\
\bar{\nu}_\mu + e &\rightarrow&  \bar{\nu}_\mu + e\nonumber \\
\bar{\nu_{e}} + e &\rightarrow&  \bar{\nu_{e}} + e\nonumber \\
e^{+}e^{-} &\rightarrow& \mu^{+}\mu^{-}\ \ \ ,\nonumber
\eea
where the last measurement involved studying the the forward-backward
asymmetry.
Taken together, these leptonic experiments implied that
$\sin^{2}\theta_{W} \approx 0.22$, in agreement with the SLAC result.
Thus,  lepton-quark universality also holds for the neutral current sector
of the $SU(2)_{L} \times
U(1)_{Y}$ theory, with a common set of couplings ($g$,$g^{\prime}$) and
mixing parameter
($\sin^{2}\theta_{W}$) governing both leptonic and semileptonic interactions.

\subsection{Atomic PV and Weak Neutral Currents}
Additional confirmation of the Standard Model structure of the weak neutral
currents
comes from atomic physics, where one looks for PV asymmetries associated
with the weak
interaction between atomic electrons and the nucleus.
In fact, one of the most precise determinations of the weak neutral
current $eq$ interaction has been performed by the Boulder group using
atomic parity-violation (APV) in
cesium\cite{boulder}. The idea behind these experiments was developed by
the Bouchiats\cite{bouchiat} in Paris in the
mid 1970's-about the same time the SLAC experiment was underway.

As in the case of PV DIS, the use of APV to get at the weak neutral current
relies on an
interference effect between the parity violating weak neutral current
atomic matrix elements and the
electromagnetic matrix elements.  However, unlike PV DIS, APV relies on a
coherent sum over the
individual electron-quark amplitudes.  The basic physics is that an atomic
electron interacts with
the nucleus by exchanging both a $\gamma$ and a $Z^0$ boson. The
probability amplitude for the
PV part of the latter is
\begin{equation}
\label{eq:apv1}
M_{PV} \sim {G_{\mu} \over 2 \sqrt{2}}[g^{e}_{A} A_{e} \cdot
<N|\sum_{q}g^{q}_{V}V_{q}|N> + g^{e}_{V} V_e\cdot
<N|g^{q}_{A}A_{q}|N>]\ \ .
\end{equation}
This amplitude, which contains two distinct terms,  causes the atomic
states of the opposite parity to
mix.  In the atomic Hamiltonian derived from this amplitude, one finds that
the first term
is dominated by the time components of the currents, whereas the second
part is dominated by the space
components.
For the first term, we have
\begin{equation}
\label{eq:apv2}
<N|\sum_{q}g^{q}_{V}V^{\mu=0}_{q}|N>=<N|\sum_{q}g_{V}^{q}q^{+}q|N>\ \ \ .
\end{equation}
Now $q^{\dag}q$ just counts the number of quarks of flavor $q$.  Thus the
matrix element (\ref{eq:apv2}) gives:
\bea
<N|\sum_{q}g^{q}_{V}V^{\mu=0}_{q}|N>&=&(2N+Z)g^{d}_{V}+(2Z+N)g^{u}_{V} \\
&=& N(2g^{d}_{V}+g^{u}_{V}) + Z(2g^{u}_{V}+g^{d}_{V}) \equiv Q_{W}\ \ \
,\nonumber
\eea
where $Q_W$ is the \lq\lq weak charge" of the nucleus.
Now
\bea
g_V^{u}&=&-1+{4 \over 3} \sin^{2} \theta_{W}\\
g_V^{d}&=&1-{8 \over 3} \sin^{2} \theta_{W}
\eea
so that the weak charges of the proton and neutron, respectively, are
\bea
Q_W^p & = & 2 g_V^u + g_V^d=1-4\sin^2\theta_W\\
Q_W^n & = & 2 g_V^d + g_V^u=-1\ \ \ .
\eea
In terms  of these quantities, the weak charge of an atomic nucleus is
\bea
Q_{W}&=&NQ^{n}_{W}+ZQ^{p}_{W}\\
&=& -N +Z(1-4\sin^{2}\theta_{W})\ \ \ .
\eea
Note that $\sin^{2}\theta_{W}\approx 0.231$ so that
$Q_W^p=1-4\sin^{2}\theta_{W} \sim 0.1 $ whereas
$Q^{n}_{W} =-1$.  Hence, in contrast to the $\gamma$ charge couplings
$Q^{p}_{EM}=1$ and
$Q^{n}_{EM}=0$, the $Z^{0}$ has a vector coupling to the neutron of
strength unity and a tiny coupling
to protons.  In short, the $\gamma$ see mostly protons, whereas the $Z^{0}$
sees mostly neutrons.
Now using  $g^{e}_{A}=1$ the first term in (\ref{eq:apv1}) becomes:
\begin{equation}
{G_{\mu} \over 2\sqrt{2}} Q_{W}e^{+}\gamma_5 e
\end{equation}
The second term is more complicated, since $<N|\sum_{q}A^{\mu}_{q}|N>$ is
not coherent
and depends on the nuclear spin(the first term does not).  Moreover,
$g^{e}_{V}=-1+4\sin^{2}\theta_{W}$
further suppresses this term.

To get some intuition into the structure of the resulting Hamiltonian,
one can take the limit of a non-relativistic electron in the field at a
point like nucleus:
\begin{equation}
\label{eq:apv3}
\hat{H}_{W}^{PV} = \hat{H}^{PV}_{W}(NSID) + \hat{H}^{PV}_{W}(NSD)
\end{equation}
where
\bea
\hat{H}^{PV}_{W}(NSID) &=& {G_{\mu} \over 4 \sqrt{2}} {1 \over m_{e}c}
Q_{W}\{\vec{\sigma} \cdot \vec{p},\delta^{3}(\vec{r})\}\\
\hat{H}^{PV}_{W}(NSD)&=& {G_{\mu} \over 4 \sqrt{2}} {1 \over m_{e}c}
(1-4\sin^{2} \theta_{W})g^{N}_{A}
\{\vec{\sigma} \cdot \vec{p},\delta^{3}(\vec{r})\}\ \ \ .
\eea
So how does an experiment which probes these interactions actually work?
The most precise result is
from APV in cesium, which has $Z=55$ and $N=78$.  The ground state of the
atom is a 6S state. The PV interaction
(\ref{eq:apv3}) mixes a bit of the 6P state into this S state. Similar
mixing occurs for the excited states.
Now, what the Colorado experiment does is apply an external electric field,
which causes mixing of the
S and P states due to the Stark effect.  In this case, one can obtain an E1
transition which is
proportional to $\vec{E}$.  The atoms are excited into the 7S state with
circularly polarized light and
the transition rate is measured.  This rate can be expressed as:\\
\begin{equation}
\Gamma(6S \rightarrow 7S) = \beta^{2} E^{2} \epsilon_{z}^{2}\left[1+K{E1_{PV} \over \beta E} {\epsilon_{x} \over \epsilon_{z}}\right]\ \ \ ,
\end{equation}
where $\beta$ is the Stark-induced amplitude, K  is a geometric factor
(dependent
on the $m$-quantum numbers),
$\epsilon_{x,z}$ are laser polarization components, and
$(E1)_{PV}$ is the PV-induced amplitude.
By reversing $\vec{E}$ or the laser polarization, one can experimentally
isolate
the interference term containing $E1_{PV}$.

It also turns out that one can experimentally isolate the effects of H(NSID),
containing $Q_{W}$, and H(NSD) by looking at sums and difference of the
rates for different E1
hyperfine transitions.  When all is said and done, one then can extract the
following quantity:
\begin{equation}
\xi Q_{W}\ \ \ ,
\end{equation}
where $\xi$ is a  quantity which depends on an atomic calculation of the PV
atomic mixing matrix
element
\begin{equation}
<P_{1/ 2}|\{\vec{\sigma} \cdot \vec{p},\delta^{3}(\vec{r})\}|s_{1/2}>\ \ \ .
\end{equation}
Taking into account the latest atomic theory computations of $\xi$, the
result of the Boulder measurement
gives\cite{sushkov}:\\
\begin{equation}
Q^{\rm exp}_{W} = -72.81 \pm 0.28({\rm exp}) \pm 0.36({\rm atomic\
theory})\ \ \ ,
\end{equation}
whereas the Standard Model prediction for $ Q_{W}$ is\cite{pdg,erler}:\\
\begin{equation}
Q^{SM}_{W} = -73.17 \pm 0.03\ \ \ .
\end{equation}

Until recently, the values of $Q_W^{\rm exp}$ and $Q_W^{SM}$ differed by up
to more than two standard deviations.
However, in the past year, atomic theorists have included have included
some rather subtle, nucleus-dependent effects in the
radiative corrections that have changed the value of $\xi$ and moved the
weak charge into agreement with the SM
prediction. Thus, taken together with the SLAC experiment and the
low-energy charged current measurements
discussed earlier, cesium APV provides a substantial vote of confidence in
the essential ingredients of the
Standard Model.

\section{Physics beyond the Standard Model}
As the previous sections have tried to illustrate, low-energy experiments
in nuclear and atomic physics
have played an important role in verifying some of the basic ingredients of
the Standard Model. Of
course, tests have been performed over a wide range of energy scales, with
some of the most decisive
having been carried out at high energy colliders. The existence of the
$W^\pm$ and $Z^0$ bosons was
discovered in collider experiments at CERN, while measurements in $e^+ e^-$
collisions at center of mass
energies $\sqrt{s}\approx 90$ GeV -- both at CERN and at SLAC -- have
tested the
properties of the neutral current sector of the Standard Model with sub-one
percent precision. Similarly,
the discovery of the top quark in $p{\bar p}$ collisions at the Tevatron
represented an important
triumph for the Standard Model, since the mass of the top quark is
consistent with what one expects
based on the $m_t$-dependence of electroweak radiative corrections to a
variety of other measured electroweak
observables. From these standpoints, the Standard Model has been an
enormously successful theory.

Nevertheless, there exist many reasons for believing that the Standard
Model is not the end of the
story.   Perhaps the most obvious is the number of independent parameters
that must be put in by hand.
As we saw earlier, the electroweak sector of the theory alone contains 17
{\em a priori} unknown parameters.
The SU(3)$_C$ sector (QCD) introduces two more, the strong coupling,
$g_{s}$, and $\theta$.  The latter
parameterizes a term in the Lagrangian:
\bea
\label{eq:thetaterm}
L^{QCD}_{\theta}& =& \theta { \alpha_{s} \over 2 \pi} G^{\mu \nu}_{a}
\tilde{G^{a}_{\mu \nu}}\\
\tilde{G_{\mu \nu}} &=& \epsilon_{\mu\nu\alpha\beta}G^{\alpha\beta}\ \ \ ,
\eea
with $G^{a}_{\mu \nu}$ being the gluon field strength tensor and
$a=1,\ldots, 8$.
Note that $L_{\theta}$ is both a pseudoscalar and odd under time reversal.
It is also even under charge conjugation,
so the interaction (\ref{eq:thetaterm}) is CP-violating. Measurements of
the electric dipole moment of the neutron and neutral atoms imply that
$\theta \leq 10^{-9}-10^{-10}$.
This seems ``un-natural'', given the size of the other parameters in the
Standard Model.  Hence the
questions:\\
What is the origin of the various parameters in the Standard Model?\\
Why do they have the values one observes them to have?\\
Why is $\theta_{QCD}$ so tiny? (This is the ``strong CP Problem'')\\
There exists already some need to go beyond the Standard Model simply to
answer these questions.
But there is even more motivation:\\
\noindent
1.\underline{Coupling Unification}\\
There exists a strongly held belief among particle physicists and cosmologists
that in the first moments of the life of the universe, all the forces of
nature were ``unified'',
that is, they all fit into a single gauge group structure whose interaction
strengths were described by
a single coupling parameter, $g_U$.  It is a remarkable idea, and an
intellectually appealing one.

\begin{figure}[h]
\psfig{figure=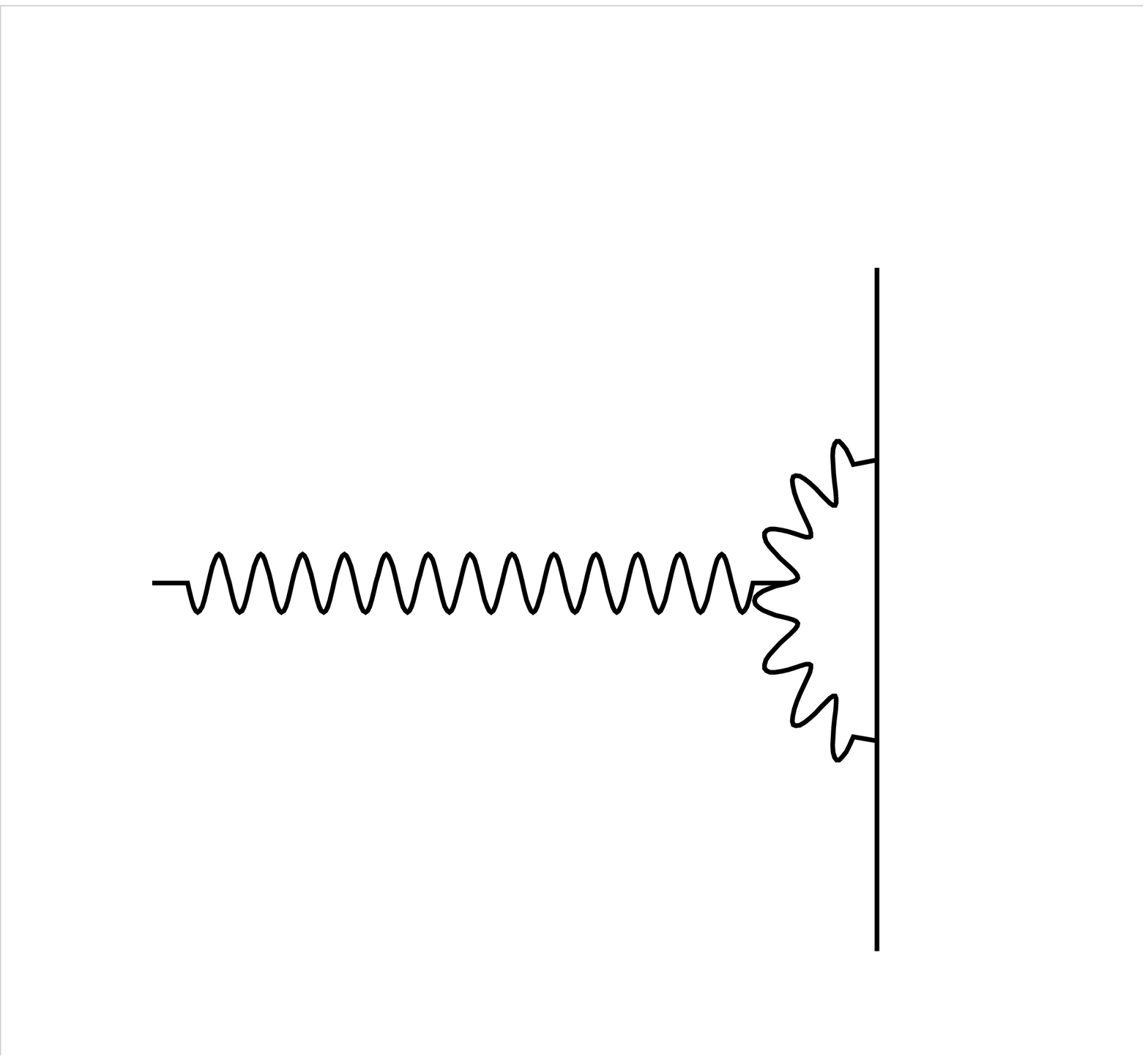,height=1.7in}
\psfig{figure=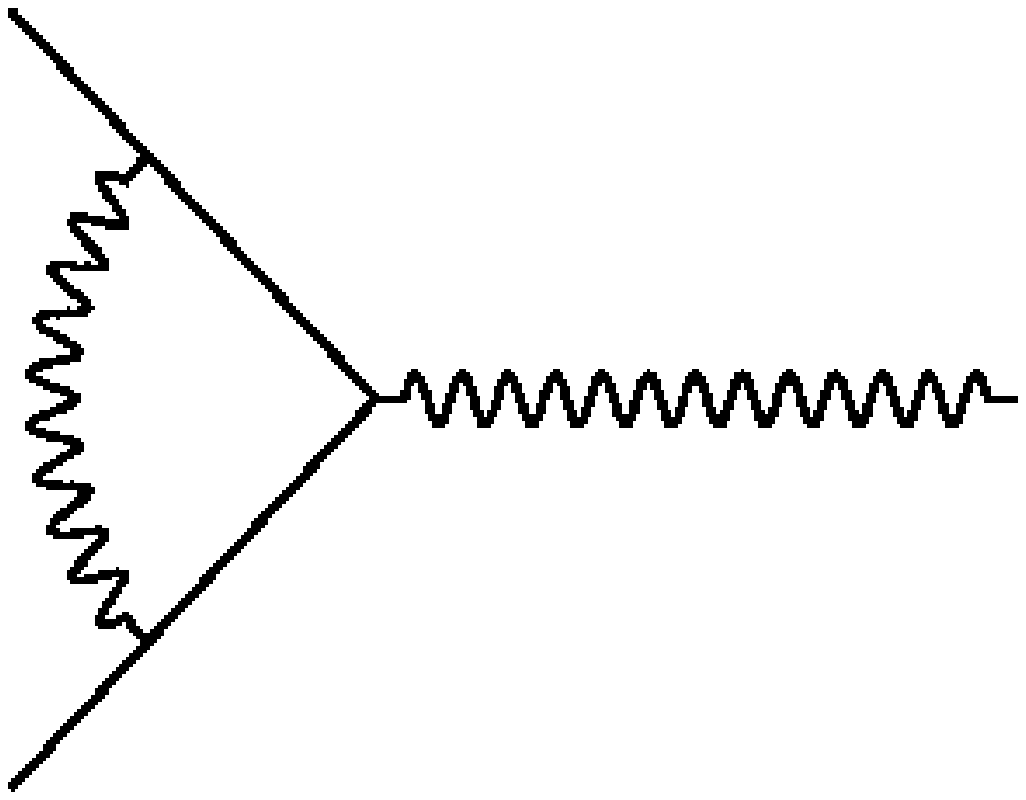,height=1.5in}
\caption{Lowest order diagrams for the renormalizing the gauge
boson-fermion vertex.}
\label{fig:diag8}
\end{figure}

\begin{figure}[h]
\psfig{figure=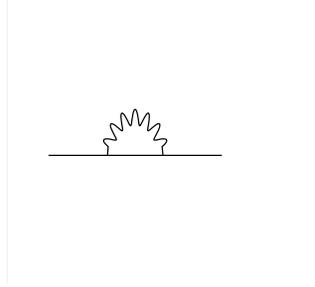,height=1.5in}
\caption{Lowest order diagrams renormalizing the fermion wavefunction.}
\label{fig:diag9}
\end{figure}

\begin{figure}[h]
\psfig{figure=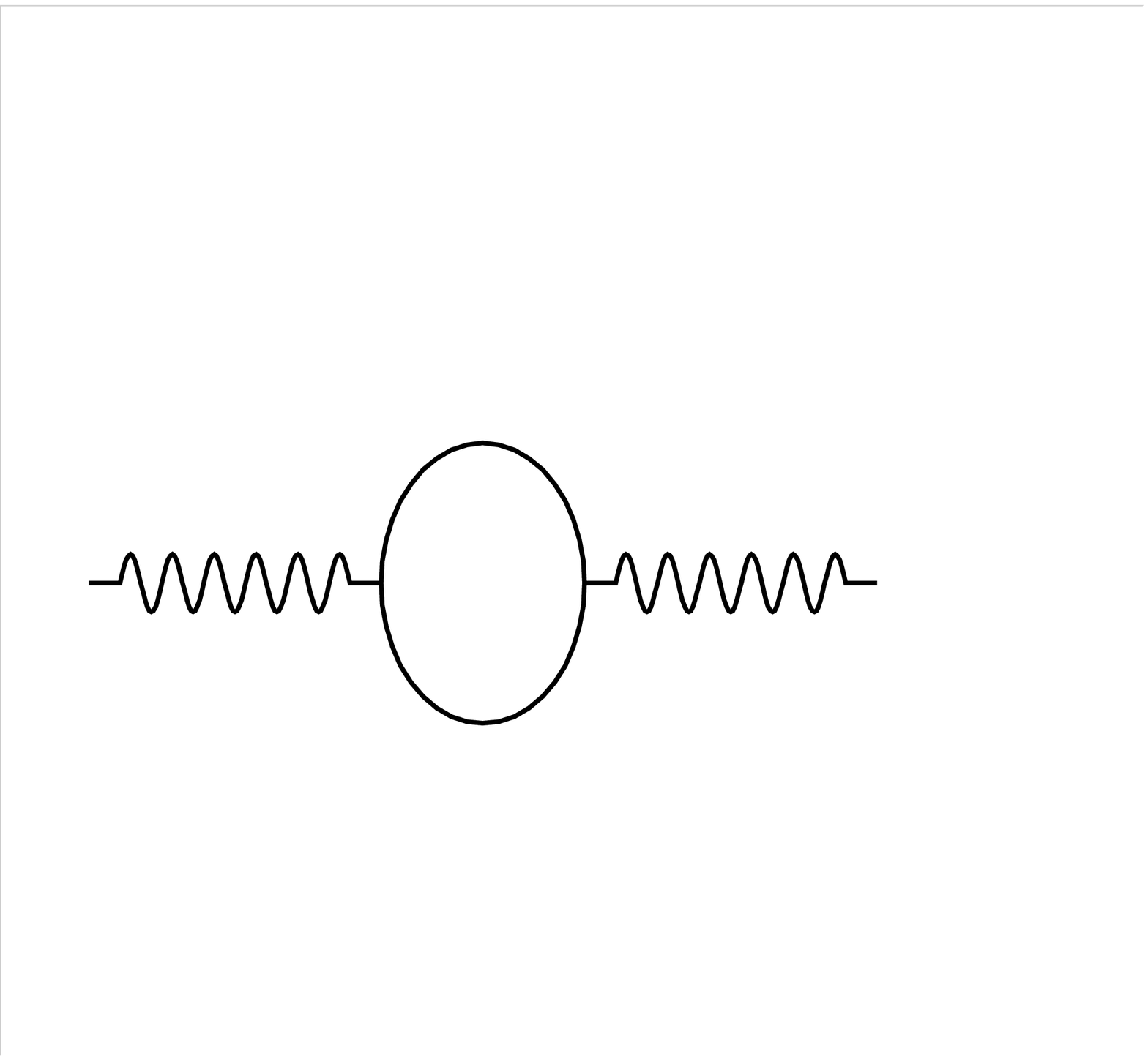,height=1.5in }
\psfig{figure=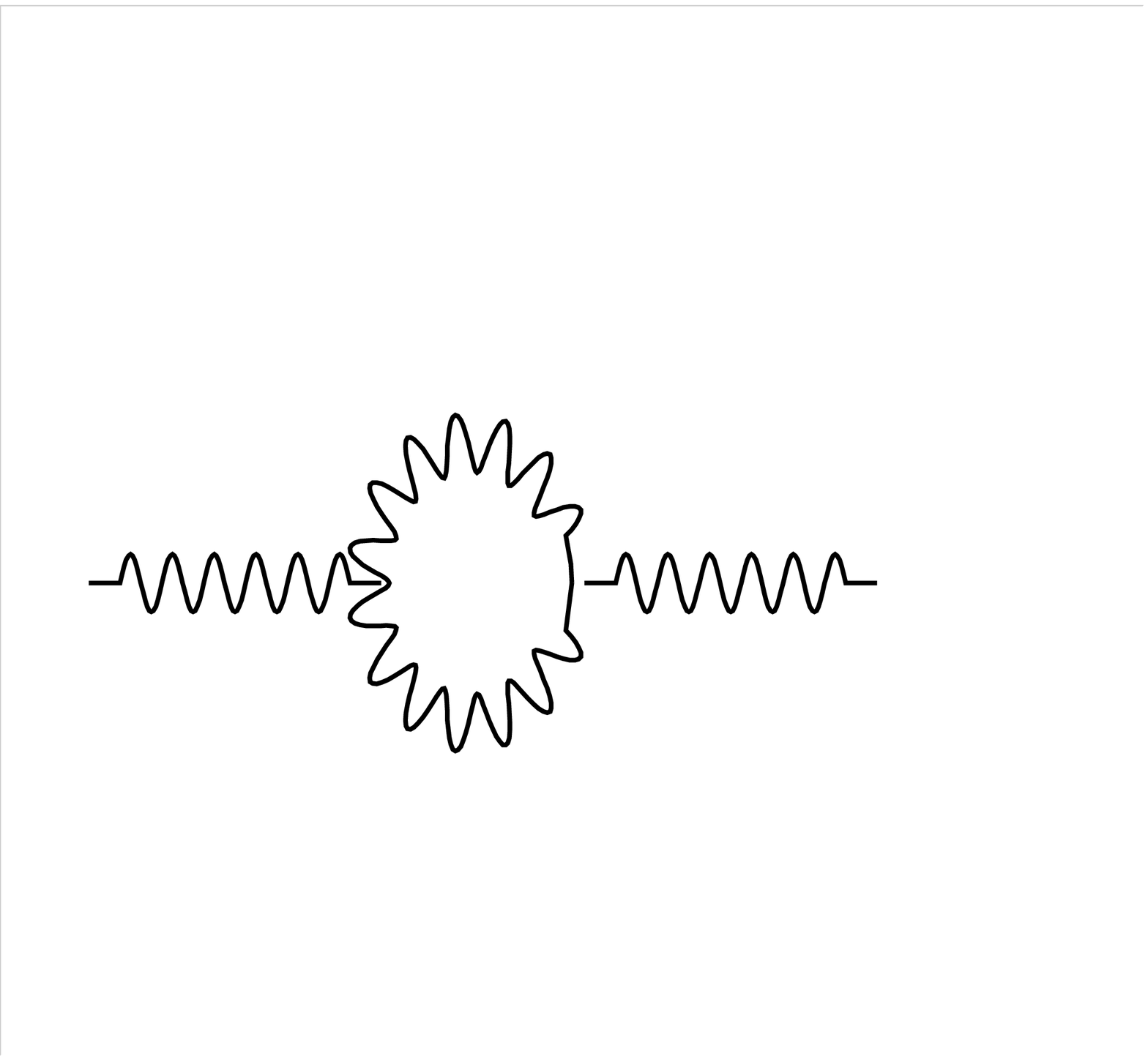,height=1.5in }
\hspace{-1cm}\vspace{2cm}\psfig{figure=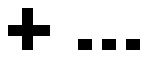,height=1.5in }
\caption{Lowest order diagrams renormalizing the gauge boson wavefunction.}
\label{fig:diag10}
\end{figure}

The scenario goes in the following manner:  as the universe cooled down,
spontaneous
symmetry breaking occurred, giving gauge bosons masses and changing the way
the interaction strength
for various forces evolved or ``ran'' down to lower energies/temperatures.
To see how this idea works,
one must work out how the couplings(which are not constants!) run with the
energy scale, $\mu$.  The
origin of this running is renormalization.  Fermion and gauge boson wave
functions, as well as
interaction vertices, get the following contributions as seen in Fig.
\ref{fig:diag8} for the vertex, Fig.
\ref{fig:diag9} for the fermion wave function and Fig. \ref{fig:diag10} for
the gauge boson wavefunction.
\bc

\ec

Each of these diagrams is infinite. This means that the theory was not
constructed correctly to begin with, and some redefinitions must be made:
\bea
\psi \rightarrow \psi_{0} = \sqrt{Z_{2}}\psi\nonumber\\
\label{eq:renorm1}
W^{a}_{\mu} \rightarrow W^{a}_{\mu 0} = \sqrt{Z_{3}}W^{a}_{\mu}\\
g \rightarrow g_{0}=g { Z_{1} \over Z_{2} \sqrt{Z_{3}}}\nonumber
\eea
The constants $Z_{i}$ are defined in such a way that when one computes the
loops with
the redefined quantities and add them to the tree level quantities, one
gets a finite answer.  The only
hitch is that in making these definitions one has to specify the energy
scale $\mu$ at which one is
working.  Hence:
\begin{equation}
Z_{i}=Z_{i}(\mu)
\end{equation}
This is because the graphs depend on $\mu$.  Now, the Lagrangian can't
depend on $\mu$;
it must hold for all scales.  Since the Lagrangian depends on the \lq\lq
bare parameters",
$\psi_{0}$, $W^{a}_{\mu 0}$, and $g_{0}$, these parameters can't depend on
$\mu$ either.  Thus, one has:
\begin{equation}
\label{eq:renorm2}
\mu {\partial \over \partial_{\mu}} g_{0}=0
\end{equation}
Comparing (\ref{eq:renorm1}) and (\ref{eq:renorm2}) one sees that $g$ must
vary with $\mu$ in order to compensate for
the $\mu$-dependence of the $Z_{i}$.  Thus, one obtains
\begin{equation}
\label{eq:renorm3}
\mu{\partial g \over \partial \mu}=\beta(g)
\end{equation}
where $\beta(g)$ depends on the group structure and fermion content of the
theory.
It dictates how $g(\mu)$ runs with $\mu$. This relationship
between $g(\mu)$ and $\beta(g)$ is known as a renormalization group equation.
One can just as well convert Eq. (\ref{eq:renorm3}) to an equation for the
running of:
\begin{equation}
\alpha_{k}(\mu)={g_{k}(\mu^{2}) \over 4 \pi}\ \ \ ,\nonumber
\end{equation}
where the subscript $k$ indicates the gauge group to which $g_k$ pertains.
The renormalization group equations for the running of $\alpha_{k}$ are ,\\
\begin{equation}
{d \over dt} \alpha^{-1}_{k} = - {b_{k} \over 2 \pi}\nonumber
\end{equation}
where $t=\ln(\mu/\mu_{0})$, $\mu_{0}$ is a reference scale, and
\bea
b_{1}={41 \over 20}, &&	\ \ \ g_{1}=\sqrt{{5 \over 3}}g^{\prime}\nonumber\\
b_{2}=-{19 \over 6}, &&	\ \ \ g_{2}=g\nonumber\\
b_{3}=-7, &&		\ \ \ g_3=g_s\nonumber
\eea

We can see that $\alpha_2({\mu})$ and  $\alpha_3({\mu})$ -- the couplings
for the non-Abelian groups -- decrease with $\mu$ whereas the U(1)$_Y$
coupling increases with $\mu$.  In
the case of QCD, this feature is known as asymptotic freedom.\\

If one were to plot the running couplings as a function of $\mu$, one would
see that the
three Standard Model couplings almost meet at a common point around
$\mu\sim 10^{16}$ GeV, but not
quite. This result is tantalizing from the
standpoint of unification.  It is one of the motivations for believing
something else is out there, as
this something else could modify the running of the couplings and produce
unification\footnote{I have not
discussed how the gravitational interaction gets incorporated into
unification. That is a separate, very difficult
problem, for which string theory may ultimately provide a solution.}.

An aside on some terminology:  One often hears reference made to the
high-energy ``desert.''
This desert is the region in $\mu$ between the weak scale,\\
\begin{equation}
M_{\rm weak} \sim 250\ \  {{\rm GeV}/ c^{2}}
\end{equation}
and the scale where one believes gravity becomes strong. The obvious
parameter which
defines this scale is Newton's gravitational constant,
\begin{equation}
G_{N}=6.67259(85) \times 10^{-11}\ m^{3}\ kg^{-1}\ s^{2}
\end{equation}
From this one may define the Planck mass:\\
\begin{equation}
M_{\rm pl}=\sqrt{\hbar c \over G_{N}} \cong 1.22 \times 10^{19}\ \  {\rm
GeV}/c^2\ \ \ .
\end{equation}
The ``desert'' then refers to the region\\
\begin{equation}
M_{\rm weak} \leq \mu \leq M_{\rm pl}\ \ \ .
\end{equation}

The interest in looking for physics beyond the Standard Model, or ``new''
physics, is really about learning what else lies in the desert.  If, in
fact, the desert is just that --
a particle physics wasteland devoid of anything new -- then one seemingly
cannot get unification.\\
\noindent
2. \underline{The hierarchy problem}\\
Suppose there does exist some new particle or particles in the desert
having  mass $m >> M_{\rm weak}$.
To be concrete, suppose the particle is a fermion. Presumably, this
particle interacts with the
Higgs boson, which is responsible for particle masses.
This interaction will affect the mass of the Higgs through higher order
diagrams.  After
renormalization, this diagram yields a finite contribution to the mass of
the Higgs:
\begin{equation}
\delta m_{H}^2= {3 |\lambda_{f}|^{2} \over 8 \pi^{2}} m_{f}^{2} ln{\mu \over m_{f}} +...
\end{equation}
Now, we know that $m_{H}$ itself must be on the order of $M_{\rm weak}$ or
below.  The reason is that one
can rewrite the Higgs potential as:
\bea
V(\Phi) &=& -\mu^{2}\Phi^{\dagger}\Phi +
\lambda(\Phi^{\dagger}\Phi)^{2}\nonumber \\
&=&-{1 \over 2}m_{H}^{2}\Phi^{\dagger}\Phi + \lambda(\Phi^{\dagger}\Phi)^{2}
\eea
with
\begin{equation}
M_{\rm weak}=v=\left({m_{H}^{2} \over 2 \lambda}\right)^{1 \over 2} \sim
250 \ \ {\rm GeV}/c^2
\end{equation}
or $m_{H} \sim \sqrt{2\lambda} \times 250\  {\rm GeV}/c^2$.
It would be unnatural for $\lambda$ (a dimensionless quantity) to be
significantly different from
unity,  so one expects that $m_{H} \sim M_{\rm weak}$.
On the other hand, if $m_{f} >> M_{\rm weak}$, one has $\delta m_{H}^2 >>
m_{H}^2 \sim M_{\rm weak}^2$!
In short, any particles which exist deep in the desert give huge
corrections to $m_{H}$,
making it unbelievable that $m_{H}$ comes out close to $M_{\rm weak}$.

To put it another way, the electroweak scale is destabilized by radiative
corrections involving heavy
particles.  It starts out at
$\sim 250\ {\rm GeV}/c^{2}$ at tree level but grows as heavy particles come
into the theory.  This
is not a desirable situation for any good theory.
The problem is known as the \lq\lq hierarchy problem":  How does the weak
scale
remain stable if the desert becomes populated?

Another aspect of the hierarchy problem is the spectrum of the Standard
Model masses themselves:
\begin{equation}
M_{\rm weak} \sim M_{W,Z} \sim m_{top} >> m_{b} >> m_{\tau} >> m_{e} >> m_{\nu}
\end{equation}
How does one explain this hierarchy of masses?
The Standard Model gives us no clue as to how to handle the hierarchy
problem.
Evidently, something new is needed.\\
\noindent
3. \underline{Quantization of Electric Charge}\\
Recall from Eq. (\ref{eq:emcharge}) that
\begin{equation}
\label{eq:chargequant}
Q=T^{L}_{3} + {Y \over 2}
\end{equation}
For any particle, $T^{L}_{3}$ is quantized in integer or half-integer units.
This follows from the algebra of SU(2):
\begin{equation}
[T_{i},T_{j}]=i\epsilon_{ijk}T_{k}\ \ \ .
\end{equation}
One the other hand, a U(1) group has no such algebra, and therefore the
eigenvalues
of the group generator Y are not restricted.  Equation
(\ref{eq:chargequant}), however implies that Y must take on
integer values for leptons and fractional values for quarks, in order to
reproduce the known fermion
charges.  This seems rather arbitrary from the standpoint of symmetry.
Why, then, is Y -- and therefore Q -- also
quantized?  The Standard Model does not motivate electromagnetic charge
quantization, but simply takes it as an
input.  The deeper origin of Q quantization is not apparent from the
Standard Model.\\
\noindent
4. \underline{Discrete Symmetry Violation}\\
By construction, the Standard Model is maximally parity-violating; it was
built to
account for observations that weak c.c. processes only involve left handed
particles (or right handed
anti-particles).  But why this mismatch between right-handedness and
left-handedness?  Again, no deeper
reason for the violation of parity is apparent from the Standard Model.

Similarly, the Standard Model allows CP-violation to creep in two places:\\
\noindent (i)  a phase in the CKM matrix\\
\noindent (ii) the QCD $\theta$ term\\
Already the $\theta$ mystery has been discussed; it is incredibly tiny for
no obvious reason in
the SM.  What about $\delta$, the CKM phase factor? Where did it come from?
In some sense, its
appearance is an artifact of the mathematics for three generations of
massive quarks.  But the reason
for the existence of three generations, and again the magnitude of the
phase factor is not explained by
the Standard Model.  It would be desirable to have answers to these
questions, but it will take some new
framework to provide them.\\

\noindent
5. \underline{Baryon Asymmetry of the Universe (BAU)}\\
Why do we observe more matter than anti-matter?  This is a problem for both
cosmology and the Standard Model. To quantify this problem, let $n_{B} =
n_{b}-n_{\bar{b}}$,
difference in the number of baryons and anti baryons per  unit volume, and
let $n_{\gamma}$ be the photon
number density at temperature T. Standard cosmology makes very accurate
predictions for the cosmological
abundance of H, $^3$He, $^4$He, $^2$H, B, and $^7$Li given that $\eta
\equiv {n_{B}/ n_{\gamma}}$
has been constant since nucleosynthesis.  The primordial abundances of
$^2$H and $^3$He imply:
\begin{equation}
\label{eq:bau}
3 \times 10^{-10} \leq \eta \leq 10 \times 10^{-10}
\end{equation}
If $\eta =0$ at the Big Bang, then standard cosmology implies that $\eta
\leq 10^{-18}$ --
much smaller than the range in (\ref{eq:bau}). Hence, the early universe
must have
$\eta\not= 0$  to explain primordial element abundances. \\
What is the connection to the Standard Model?  It was provided by Sakharov,
who
pointed out that a non-vanishing $\eta$ may exist in the early universe if:\\
(i) Baryon number is violated
\medskip
(ii) Both C and CP are violated
\medskip
(iii) At some point, there has bee a departure from thermal equilibrium.
\medskip
\noindent These are known as the Sakharov criteria\cite{sakharov}. The
reason for (i) is clear.  The violation of both
C and CP is needed so that:\\
\begin{equation}
\nonumber
\Gamma({\rm baryon\ production}) \not= \Gamma ({\rm anti-baryon\
production})\ \ \ .
\end{equation}
The third Sakharov criterion is needed to get a non-zero thermal average of B.
Now, it is known that in the Standard Model, B+L is broken by instanton
effects.
Similarly, the Standard Model has maximal C-violation. Consider, for
example, the
charged current interaction
\begin{equation}
\nonumber
\bar{u}\gamma^{\mu}(1-\gamma_{5})d\ W_{\mu}^{-} \ \ \ .
\end{equation}
The axial vector part of this interaction is C-odd because
\bea
\nonumber
\bar{u}\gamma^{\mu}\gamma_{5}d &{\rightarrow\over {C}} & +
\bar{u}\gamma^{\mu}\gamma_{5}d \\
\nonumber
W_{\mu} &{\rightarrow\over {C}} &-W_{\mu} \ \ \ .
\eea
CP violation enters via the CKM phase factor $\delta$ as well as the
$\theta$ parameter.
As noted earlier, $\theta$ is incredibly tiny -- far too small to provide
the necessary amount
of CP-violation for the baryon asymmetry. Similarly, the
magnitude of $\delta$'s  contribution to the baryon/antibaryon asymmetry is
significantly smaller
than needed to produce the required value of $\eta$.\footnote{The magnitude
of this contribution depends not only
on $\delta$ but also on the other angles appearing in the CKM matrix.}
Thus, if one is going to live within standard cosmology, one needs
additional sources of large CP-violation beyond the
Standard Model to explain BAU.\\

To summarize, despite the triumphant successes of the Standard Model,
there exist conceptual motivations for believing that there is something
more, that the high energy
desert is not so barren after all.

One of the goals in experimental high energy physics -- as well as in
precision
low energy electroweak experiments -- is to go looking for new physics.  In
fact, the results of
these searches can constrain the new physics scenarios people have
invented, or at least dictate what
some of the parameters in these scenarios must be. Some of the most popular
such scenarios include
supersymmetry (SUSY), extended gauge symmetry, and extra dimensions. The
appeal of SUSY is that it provides a natural
solution to the hierarchy problem, produces unification of gauge couplings,
and contains new CP-violating effects
that could help produce a sufficiently large BAU. Extended gauge theories,
on the other hand, can also produce gauge
unification, provide a natural mechanism for electric charge quantization,
and and can account for the violation of
parity invariance in low-energy weak interactions. Finally, the idea that
we live in more than four spacetime
dimensions -- which has been motivated by string theory -- gives an
alternative solution to the hierarchy problem
than contained in SUSY. The implications of this paradigm for the
phenomenology of electroweak interactions is now a
lively area of research in particle physics.

Given the scope of these lectures, I do not have the time and space to
discuss these scenarios in any depth. At the
very least, however, I hope to have provoked your curiosity and motivation
for learning more about them. We should
always keep in mind, that however one seeks to
extend the Standard Model, one should take care to respect the basic
ingredients of the Standard Model and its
phenomenological successes:
\bc
Gauge symmetry\\
Universality\\
V-A dominance\\
Quark Mixing\\
neutral currents and $\sin^{2}\theta_{W}$\\
conserved vector currents\\
etc.\\
\ec
Any deviations from the Standard Model predictions based on these
ideas must be small, and any new physics scenario must explain why it only
produces small deviations
in a natural way.





\begin{acknowledgements}
This work was supported in part under U.S. Department of Energy contracts
DE-FG02-00ER41146, DE-FG03-02ER41215, DE-FG03-88ER40397, DE-FG03-00ER41132 and
NSF Award PHY-0071856.
\end{acknowledgements}


\end{document}